# Longitudinal variations of positive dayside ionospheric storms related to recurrent geomagnetic storms

A. V. Dmitriev[1,2], C.-M. Huang[1], P. S. Brahmanandam[1,3], L. C. Chang[1], K.-T. Chen[1], L.-C. Tsai[4]

[1]*Institute of Space Science, National Central University, Chung-Li, Taiwan*

[2]*Skobeltsyn Institute of Nuclear Physics Moscow State University, Moscow, Russia*

[3]*Dept. of ECE, K L University, Vaddeswaram, AP, India*

[4]*Center for Space and Remote Sensing Research, National Central University, Chung-Li, Taiwan*

Short title: RECURRENT IONOSPHERIC STORMS

A. V. Dmitriev, Institute of Space Science, National Central University, Chung-Li, 320, Taiwan, also at D.V. Skobeltsyn Institute of Nuclear Physics, Moscow State University, Russia (e-mail: dalex@jupiter.ss.ncu.edu.tw)

C.-M. Huang, Institute of Space Science, National Central University, Chung-Li, 320, Taiwan (e-mail: cmh@jupiter.ss.ncu.edu.tw)

P. S. Brahmanandam, Institute of Space Science, National Central University, Chung-Li, 320, Taiwan (e-mail: anand.potula@gmail.com) also at Dept. of ECE, K L University, Vaddeswaram, India.

L. C. Chang, Institute of Space Science, National Central University, Chung-Li, 320, Taiwan (e-mail: loren@jupiter.ss.ncu.edu.tw)

K.-T. Chen, Institute of Space Science, National Central University, Chung-Li, 320, Taiwan (e-mail: bestman8556@gmail.com)

L.-C. Tsai, Center for Space and Remote Sensing Research, National Central University, Chung-Li, Taiwan (e-mail: lctsai@csrsr.ncu.edu.tw)

**Abstract**

We have performed an analysis of case events and statistics of positive ionospheric storms in the dayside region of the equatorial ionization anomaly during recurrent geomagnetic storms (RGSs), which dominate in geomagnetic and ionospheric conditions on the declining phase of solar activity in 2004 to 2008. It is shown that total electron content (TEC) has a tendency to minimize before the beginning of RGSs and to peak 3 to 4 days after, i.e. on the RGS recovery phase produced by high-intensity long-duration continuous auroral activity. The maximum of TEC coincides with the maximum of solar wind velocity within high-speed solar wind streams. An analysis of electron content vertical profiles, derived from two independent methods using ionosondes and COSMIC/FORMOSAT-3 radio occultation, showed that in the maximum of an ionospheric storm on 28 March 2008, the F2 layer thickens, NmF2 increases by ~50% and hmF2 elevates by a few tens of kilometers. The response of positive ionospheric storms to solar, heliospheric and geomagnetic drivers reveals a prominent longitudinal asymmetry. In the longitudinal range from –90° to 90°, the solar illumination plays a major role, and in the range from 90° to –120°, the influence of heliospheric and geomagnetic drivers becomes significant. The highest correlations of the TEC enhancements with the heliospheric and geomagnetic drivers were found during December - February period ($r$ increased from ~ 0.3 to ~0.5). We speculate that the dynamics controlling this might result from an effect of solar zenith angle, storm-time effects of thermospheric $\Sigma O/N_2$ enhancement, and penetrating electric fields of interplanetary and magnetospheric origin.

## 1. Introduction

In recent years, a great deal of interest has been directed to recurrent geomagnetic storms (RGSs) and their ionospheric and thermospheric effects [*e.g. Lei et al.*, 2008a;b;c; 2011; *Mlynczak et al.*, 2008; *Thayer et al.*, 2008; *Denton et al.*, 2009; *Sojka et al.*, 2009; *Verkhoglyadova et al.*, 2011; *Liu et al.*, 2012a]. Comprehensive studies of the ionospheric day-to-day variability reveal that geomagnetic and meteorological sources of the variations in the F2 layer ionization are comparable while the solar radiation source is a minor contributor [*Forbes et al.*, 2000; *Rishbeth and Mendillo*, 2001]. The regularity of interplanetary driving and thermospheric/ionospheric responses suggests a possibility to develop a reliable 27-day prediction of ionospheric disturbances.

RGSs result from interaction of the magnetosphere with a complex interplanetary structure consisting of high-speed streams (HSSs) of the solar wind preceded by corotating interaction regions (CIR) formed between the fast and slow solar wind streams [*Burlaga and Lepping*, 1977; *Tsurutani et al.*, 1995; 2011a]. During rising and declining phases of the solar cycle, RGSs occur almost every 7 to 13 days because of the draped heliospheric current sheet related to longitudinal distribution of active regions and coronal holes on the Sun. The most dramatic geomagnetic response to the HSSs are chains of consecutive substorms caused by the southward components of large-amplitude Alfven waves within the body of HSSs. This auroral activity has been named high-intensity long-duration continuous AE activity (HILDCAAs) [*Tsurutani et al.*, 1995].

The RGSs are typically not very intense [*Richardson et al.*, 2006]. They peak at *Dst* ~ -40 nT but have a prominent seasonal variation of ~40 nT which is ordered by the spring and fall equinoxes. The substorm activity is generally most intense near the peak of HSS where the Alfven wave amplitudes are greatest, and it decreases with decreasing wave amplitudes and solar wind speed. Hence, the RGSs have long-lasting main and recovery phases. The main phase is accompanied by relatively high solar wind dynamic pressure and rapidly varying strong IMF *Bz*, associated with the CIR. This phase is characterized by intense particle precipitation in the auroral region and from the radiation belt to lower latitudes. The recovery phase is accompanied by HSSs and related HILDCAAs, which contribute the fresh and sporadic injection of substorm energy leading to unusually long storm recovery phases (up to ~10 days) as noted in *Dst*. This phase is also often accompanied by relatively low solar wind density and weakened alternating IMF *Bz*.

There are a number of differences between the RGSs and CME-driven geomagnetic storms and their ionospheric manifestations [*Borovsky and Denton*, 2006; *Liu et al.*, 2012]. The main difference consists in the penetration of the interplanetary electric field (IEF) in the magnetosphere. In contrast to prompt (time scale of minutes) penetration for CME-driven storms [e.g. *Guo et al.*, 2011], during CIRs, the daytime ionospheric responses to IEF changes were found to occur within ~1 hour [*Koga et al.*, 2011]. *Burns et al.* [2012] suggested that while CIRs are substantial factor affecting the thermosphere-ionosphere response, the level and duration of the continuing forcing during the HSSs is also important. This continuing forcing leads to extended periods of storm effects in the ionospheric electron content and enhanced neutral densities for many days after the end of the CIR phase. Thus, although the magnitudes of RGSs are considerably smaller than those of CME-driven storms, their cumulative effects on the thermosphere and ionosphere are comparable [e.g. *Turner et al.*, 2009].

A distinct global response of the ionosphere to RGSs was reported in a number of papers [e.g. *Denton et al.*, 2009; *Ram et al.*, 2010; *Pedatella et al.*, 2010; *Pedatella and Forbes*, 2011; *Verkhoglyadova et al.*, 2011; *Liu et al.*, 2012a]. On average, the electron content in the low-latitude ionosphere increases during the CIR-related main phase of RGSs. The virtual height of the F2 layer peak (hmF2) on the dayside is found to increase by a few tens of km. During the long-lasting recovery phase related to HSS, the electron content decreases but persists higher than the pre-storm level. *Ram et al.* [2010] pointed out that the RGSs are characterized by a strong altitude and latitude dependence of electron density perturbations, which can be explained by such physical processes as photoionization-chemistry, particle precipitation, and dynamic and diffusion transport. The disturbance dynamo electric field (DDEF) [*Blanc and Richmond*, 1980; *Huang et al.*, 2005] is also considered to be an important mechanism for driving the electrodynamic response of the dayside ionosphere to recurrent geomagnetic activity related to high-speed solar wind streams [*Pedatella et al.*, 2010; *Pedatella and Forbes*, 2011; *Liu et al.*, 2012].

Fast (within a few hours), global, and continuous ionospheric responses to RGSs were reported by a number of authors [e.g. *Ram et al.*, 2010; *Verkhoglyadova et al.* 2011]. The largest ionospheric variations were found at low latitudes. It was concluded that CIRs/HSSs were external drivers for both thermospheric and ionospheric phenomena. A global thermospheric heating was revealed during RGSs. For instance, *Sojka et al.* [2009] found a significant increase of the ion temperature in the high-latitude ionosphere during CIR intervals. Prominent increases of low-latitude thermospheric density and ΣO/N2 during RGSs were also found in a number of studies [*Lei et al.*, 2008a; 2011; *Crowley et al.*, 2008; *Burke et al.*, 2010; *Liu et al.*, 2012b]. *Burke et al.* [2010] have shown that a model used to estimate exospheric temperature changes during large CME-driven storms overpredicts the thermospheric heating during RGSs. However, large-amplitude Alfvén waves in the interiors of HSSs generate regularly observed increases in the exospheric temperature. *Deng et al.* [2011] concluded that the energy transfer process into the upper atmosphere associated with HSSs is a combination of Joule heating and particle precipitation at high latitudes, while Joule heating plays a dominant role.

An important manifestation of efficient heliospheric-magnetospheric-ionospheric-thermospheric coupling occurring during RGSs is a 9-day periodicity in variations of the ionospheric and thermospheric parameters [*Crowley et al.*, 2008; *Lei et al.*, 2008a;b;c; *Mlynczak et al.*, 2008; *Thayer et al.*, 2008; *Chang et al.*, 2009; *Pedatella et al.*, 2010; *Ram et al.*, 2010; *Liu et al.*, 2010a; 2012a]. On average, at low-middle latitudes, the 9-day oscillations in thermospheric density, temperature and such ionospheric parameters as electron density in the maximum of F2 layer (NmF2), hmF2 and the thickness of F2 layer (HT) are in phase with those in geomagnetic activity.

However, in the equatorial anomaly region, the ionosphere shows more complicated day-to-day variability. *Wang et al.* [2011] found that low-latitude stations, located at different longitudes, detected ionospheric variations with different periodicities of 11 and 16 – 21 days, though the 9-day periodicity was dominant. The difference indicates a longitudinal inhomogeneity in the ionospheric day-to-day variability related to geomagnetic variations as well as to other sources.

It can be seen that the pattern of the ionospheric response to the recurrent geomagnetic activity is very complex and quite different from the well-developed model of “standard” storms proposed by *Fuller-Rowell et al.* [1994]. However, the cumulative ionospheric effect of RGSs and CME-driven storms can be comparable. The complexity of temporal and spatial dynamics of the ionosphere during RGSs results from strong and fast variations of external drivers such as IEF and solar wind dynamic pressure, together with intense particle precipitations in the magnetosphere that results in strong variations of prompt penetration electric field (PPEF), disturbance dynamo electric field (DDEF), thermospheric density and chemical composition [e.g. *Wang et al.*, 2011; *Burns et al.*, 2012; *Liu et al.*, 2012a]. Past studies on RGS-related disturbances in the thermosphere and ionosphere focused mainly on the zonally symmetric component that assumes an identical response at all longitudes. In addition, statistical studies of RGS-related ionospheric/thermospheric disturbances did not distinguish between the CIR-related main phase and HSS-related maximum and recovery phase of RGSs.

In the present paper, we investigate longitudinal variations of the ionization enhancements in the low-latitude dayside ionosphere (so-called positive ionospheric storms) during different phases of RGSs. The organization of present paper is as follows: Statistical study of positive ionospheric storms is presented in Section 2. Vertical profiles of enhanced electron content are investigated in Section 3. The results are discussed in Section 4. The conclusions are given in Section 5.

## 2. Positive ionospheric storms

We study positive ionospheric storms that occurred from 2004 to 2008 during the declining phase of solar activity. As shown in Figure 1, this interval is characterized by relatively low levels of the solar radiation represented by the solar radio flux with wavelength of 10.7 cm (F10.7 index). The F10.7 index can be considered as a proxy of the solar radiation controlling the dayside ionospheric ionization [*Ozguc et al.*, 2008; *A et al.*, 2012]. During declining and minimum phases as well as during rising phases of solar cycle, the heliospheric and geomagnetic conditions are dominated, respectively, by CIR/HSSs and by RGSs, whose amplitude, measured as a minimum *Dst* variation, does not usually exceed several tens of nT [e.g. *Bothmer et al.*, 2004]. We do not consider the year of 2009 when a deep minimum occurred in solar, heliospheric, geomagnetic and ionospheric/thermospheric activity [*Tsurutani et al.*, 2011b; *Verkhoglyadova et al.*, 2013].

**2.1. Global ionospheric maps**

The long time scales of RGSs make it possible to study ionospheric effects of these storms with 2-hour resolution. This time resolution is widely used for a global ionosphere mapping such 2-D global ionospheric maps (GIMs) of vertical total electron content (VTEC) derived from the global positioning system (GPS) network [*Rebischung et al.*, 2012; http://aiuws.unibe.ch/ionosphere/]. GIMs are generated using data acquired from about 200 ground-based receivers of the radio-signals from GPS and GLONASS satellite constellations and analyzed by the International Global Navigation Satellite Systems Service (IGS) and other institutions [http://igscb.jpl.nasa.gov/network/refframe.html]. The reliability of GIMs was investigated in a number of studies [e.g. *Hernandez-Pajares et al.*, 2009; *Jee et al.*, 2010]. It was shown that on the whole, GIMs were largely able to reproduce the spatial and temporal variations of the global ionosphere as well as annual and semiannual variations and solar cycle variations.

Table 1 shows number of IGS receivers used for construction of GIM at low- to mid-latitudes in different longitudinal ranges. Above the Pacific Ocean (longitudes from 150° to 180° and from –180° to -120°), the number of receivers is less than that above continents. However, each longitudinal range is represented by at least 4 receivers located in both Southern and Northern hemispheres at low- to mid-latitudes that allows for measuring a region of the equatorial ionization anomaly (EIA).

An example of GIMs is shown in Figure 2. The maps, constructed during a recurrent magnetic storm on 27 and 29 March 2008, are compared with a quiet day on 25 March 2008. Solar, heliospheric and geomagnetic conditions during the storm are presented in Figure 3. Namely, March 27 and 29 correspond, respectively, to the maximum and recovery phase of the storm. The residual VTEC reveals prominent positive ionospheric storms, with enhancements up to >20 TECU occurring at low latitudes in the postnoon and evening sectors in Indochina, Pacific, and American regions. It is important to note that the largest enhancements of VTEC are observed in the crest of EIA. Hence, it is possible to analyze the positive ionospheric storms by studying the maximum VTEC in the EIA region.

The 2-hour time step of GIMs is equivalent to a 30° step in longitude. That allows for dividing the map onto 12 longitudinal ranges: from -180° to -150°, from -150° to -120° … from 150° to 180°. We can, thus, construct a portion of GIM every two hours and for each longitudinal range. Spatial VTEC distribution varies with local time and, hence, with UT such that higher (lower) VTEC appear on the dayside (nightside). For each day and for each longitudinal range, we determine a daily maximum VTEC (maxVTEC).

Examples of variations of VTEC distributions and maxVTEC within 27-day intervals in various longitudinal ranges are presented in Figures 3 and 4, respectively, during spring equinox (from 20 March to 15 April 2008, a part of Whole Heliospheric Interval [e.g. *Verkhoglyadova et al.*, 2011]) and near the winter solstice (from 1 January to 3 February 2007). The interval of 27 days is approximately equal to a period of solar rotation in the Earth's frame that is a prime source of the recurrent variations. Figures also show time profiles of daily solar radiation (F10.7), and 1-min data of solar wind velocity (*Vsw*), density (*D*), IMF *B*z in GSM coordinates, and 1-hour geomagnetic indices *AE*, *Dst* and 3-hour *Kp*. Solar wind parameters were acquired from ACE and Wind upstream monitors. The Wind monitor was used when the ACE data was poor due to numerous data gaps. For the days when neither ACE nor Wind data were available, such as 26 March 2008 in Figure 3, we eliminated the solar wind data from consideration.

During 27-day periods, we observe 2 or 3 intervals of co-rotating CIR-HSS structures, characterized by high solar wind speeds of >400 km/s. The total duration of these structures varies from 4 to 10 days. The co-rotating streams result in recurrent geomagnetic storms of similar duration and repeating time. The onset of RGS is determined as an abrupt increase of geomagnetic indices *AE*, *Kp* and *Dst*. The increase results from magnification of southward IMF in compressed solar wind of the CIR region. The time interval between the onsets varies from 7 to 13 days. The storms are characterized by a high level of the auroral activity (large *AE* index) as well as by global geomagnetic disturbances revealed in enhanced *Kp* and *Dst* indices (with minimum *Dst* of about –40 nT).

As one can see in Figure 3, maxVTEC occurs mainly in the crest of the EIA. On March 24 – 25, maxVTEC increased with the solar radiation index F10.7. The increase was stronger in the longitudinal range from –120° to –90° and weaker in the ranges from 90° to 120° and from –30° to 0°. At the same time, maxVTEC also enhanced during recurrent magnetic storms that occurred on March 26 – April 2 (8-day duration) and April 4 – 14 (11-day duration). Note that the former storm overlapped with *F*10.7-related increase of VTEC. The VTEC enhancements were mostly prominent in the longitudinal range from –120° to –90°. It is interesting to note that maxVTEC peaks several days after the onset of RGSs. Namely, the maximum values of maxVTEC occurred on March 27 – 28 (2 – 3 days after the RGS onset) and on April 5 – 8 (1 – 4 days after the RGS onset). In the range from 90° to 120°, an additional prominent maximum of maxVTEC occurred on April 10, i.e. 6 days after the RGS onset.

It is interesting to note that the CIR-related onsets of RGS (March 26 – 27 and April 4 – 5) were accompanied by suppressions of maxVTEC. Most prominent suppressions were observed in the longitudinal range from –120° to –90°. At ~22 UT on April 4, the suppression was accompanied by the northward IMF. On the other hand, another prominent suppression can be found at ~14 UT on April 4 in the longitudinal range from –30° to 0°. At that time, the IMF was southward and large.

During the late recovery phase of RGSs, maxVTEC was decreasing at all longitudes. The decrease was not gradual but sustained a day-to-day variability with local minima and maxima occurred at different longitudes during different days. Very low maxVTEC in the EIA region was observed at all longitudes during a few days between the recurrent storms (March 24, April 2-3 and 14). Those days are characterized not only by very low geomagnetic activity but also by quiet interplanetary conditions characterized by low solar wind speed and weak IMF $B$z variations.

A similar pattern of the VTEC dynamics can be found during winter solstice (Figure 4). Note that maxVTEC occurs mainly in the southern crest of the EIA that corresponds to better illumination of the Southern hemisphere. From January 8 to February 4, we distinguish two RGSs: January 14 – 26 (13-day duration) and January 29 – February 4 (7-day duration). The former (latter) storm was accompanied by a decrease (increase) of $F$10.7. Quiet days in the ionosphere can be identified via low maxVTEC on January 13 and 27-28. They were characterized rather by low solar wind speed and weak IMF $B$z variations while geomagnetic activity was slightly disturbed.

Suppressions of maxVTEC can be seen during CIR intervals on January 14-15 in longitudinal ranges from 0° to 30° and from 120° to 150° and on January 30 in all longitudinal ranges. Note that on January 14, northward IMF accompanied the suppression in the range from 0° to 30° while the suppression in the range from 120° to 150° occurred on January 15 under southward IMF. January 30 was characterized by highly variable IMF. Hence, it seems the suppression is driven not only by the IMF orientation.

Enhancements of maxVTEC are observed at all longitudes on January 16 – 17 (2 – 3 days after the RGS onset) and on January 31 – February 2 (2 - 4 days after the RGS onset). It is important to point out that the RGS-driven variation of maxVTEC is comparable with and, thus, can be hidden by the variation related to $F10.7$. Therefore, we have to separate these two drivers carefully and such efforts have been made in this research.

Hence, the maximal VTEC occurs usually 2 to 4 days after the RGS onset and sometimes even on the late stage of the recovery phase of RGSs. Most prominent enhancements of maxVTEC were rather related to HSSs. The amplitude of maxVTEC variations during RGSs can be estimated as being in the range of tens of TECU. In addition, we have found a difference of the order of ~ 10 TECU in the ionospheric response to RGSs in different longitudinal sectors. Namely, the variations of maxVTEC in the longitudinal ranges from –150° to –90° and from 90° to 150° are much stronger than those in the range from –30° to 30°.

It is important to point out that the ionosphere and geomagnetic activity is disturbed by RGSs during most of time. There are only a few quiet days in the ionosphere per 27-day solar rotation period. Those days are characterized by both low geomagnetic activity and quiet solar wind. Similar patterns can be found for all other 27-day intervals in 2004 – 2008. Therefore, RGS-related ionospheric effects contribute a significant portion of the statistics.

### 2.2. Superposed epoch analysis of RGS-related maxVTEC variations

In order to study ionospheric effects of recurrent geomagnetic storms, we consider only weak and moderate magnetic storms with $Dst$ > -70 nT. As one can see in Figure 1, strong magnetic storms with peak $Dst$ < -70 nT produce only a small portion of the statistics. Superposed epoch analysis is based on 185 RGSs occurred from 2004 to 2008. The time is calculated from the onset of RMS (day

= 0). The negative and positive time corresponds to periods, respectively, before and after the storm onset.

A distribution of local time of RGS onsets is presented in Figure 5. We distinguish four local time sectors: midnight (LT = 21 – 3), morning (LT = 3 – 9), noon (LT = 9 – 15), and evening (LT = 15 – 21). It can be clearly seen that the occurrence probability of onsets does not practically depend on the local time. An excess (deficiency) of the onsets in the noon (evening) sectors does not exceed 2 standard errors. Hence, longitudinal differences in the ionospheric response (if any) should weakly depend on the differences in local time of RGS onsets.

The superposed epoch analysis is performed for daily values. Note that for each storm, the time starts (day = 0) from the onset such that the analyzed days are different from the calendar days. Within 10-day time intervals starting 2 days before the onset and lasting 8 days after (if possible), we calculate for each day a minimum of *Dst* and *Bz*, maximum of *Vsw*, *D*, *Kp*, and *AE* as well as maxVTEC in various longitudinal sectors. Note that RGSs with duration less than 8 days contribute to the statistics of the first few days only.

Figure 6 demonstrates examples of the superposed epoch analysis applied for minimum *Dst* and maxVTEC in longitudinal ranges from 0° to 30° and from 120° to 150°. These ranges are well covered by IGS receivers (see Table 1). For comparative analysis of maxVTEC variations, we subtract mean values of maxVTEC calculated upon all statistics in different longitudinal ranges (see Figure 7) such that maxVTEC varies around zero. Recurrent storms are characterized by a prominent day-to-day geomagnetic and ionospheric variability, which is exhibited by jumps of *Dst* and maxVTEC. This variability is clearly seen in Figures 3 and 4. Variations of maxVTEC in the longitudinal range from 120° to 150° exceed 10 TECU quite often while variations in the range from 0° to 30° are mainly below 10 TECU. Variations of median maxVTEC, calculated upon all RGSs, have amplitudes, respectively, of 3.2 and 2.4 TECU that corresponds to ~30% difference. Note that this difference between medians is significant because it results from large statistics of 185 RGSs.

In Figure 6, variations of median *Dst* and median maxVTEC exhibit a certain temporal pattern. Namely, largest values of *Dst* (~ -10 nT) are revealed before and on the first day of RGSs (day = -2 to 0). The median *Dst* reaches minimum of about -27 nT on the second day of the storm (day = 1) and then *Dst* restores gradually. During RGSs, day-to-day variability of *Dst* is produced by two competitive drivers: solar wind pressure *P*d and IMF *B*z. Positive variations in *Dst* are produced by northward turning IMF (*B*z > 0) as well as by enhancements in the *P*d (solar wind density and/or velocity) resulting in magnetospheric compression and intensification of Chapman-Ferraro current at the magnetopause [e.g. *O'Brien and McPherron*, 2002]. Increases of southward IMF (*B*z < 0) lead to negative *Dst* variations. It seems that the small negative *Dst* ~ -10 nT during the RGS onset results from high *P*d whose geomagnetic effect is stronger than the effect of southward IMF. The latter can be strong but its duration is usually short that does not allow large negative *Dst* variation.

The variation of median maxVTEC follows the *Dst* variation only partially. A prominent decrease of median maxVTEC is observed right before the onset (day = -1). Then maxVTEC increases gradually during 3 days up to the maximum values (day = 2) and decreases during the rest of time. Note that the maximum of median maxVTEC occurs one day after the minimum of *Dst* that might be related to HILDCAAs. The onset of RGS can be accompanied by both suppressed and enhanced maxVTEC. In the range from 0° to 30°, negative variations occur more often such that the variation of median maxVTEC at day = 0 is negative (-0.3 TECU).

Figure 7 shows longitudinal changes in maxVTEC variations. The amplitude of variations (a difference between maximum and minimum) of the median maxVTEC increases by ~2 times from ~ 2 TECU in the longitudinal range –60° - 120° to ~4 TECU in the range 120° - -60° (i.e. from 120° to 180° and from –180° to –60°). This difference can be considered to be significant because it was obtained from large statistics of 185 RGSs. The variations of median maxVTEC at day = 0 are negative practically at all longitudes that indicates a predominant suppression of maxVTEC during the RGS onsets.

For each longitude range, we also estimated an average enhancement of maxVTEC by averaging the maximal values of maxVTEC observed during each RGS. The average enhancement varies with longitude substantially (~60%) from ~5 TECU in the longitudinal range from -60° to 60° to ~8 TECU in the range from longitudes from 60° to -60°. Hence, we can conclude that the longitudinal regions from -60° to 60° and from 120° to –60° are characterized, respectively, by lowest and highest variability of maxVTEC. Note that both regions contain several longitudinal ranges with good coverage by IGS receivers that support the validity of our results.

We have to point out that the present pattern of longitudinal variations in maxVTEC is quite different from the pattern of LT variation in RGS onsets presented in Figure 5. Hence, the weak LT variation of RGSs has a minor contribution to the observed longitudinal variation of maxVTEC, which would be driven by other significant effects.

In Figure 7 we also present mean values of maxVTEC calculated for each longitudinal range as average of maxVTEC upon whole statistics, including quiet days and 185 RGSs. The mean maxVTEC demonstrates a prominent longitudinal variability of ~15% with minimum around 41 TECU in the longitudinal range from –60° to 60° and maximum around 47 TECU in the ranges from –180° to –120° and from 90° to 120°. It is important to point out that similar longitudinal variation with amplitude of a few tens of percent was revealed by *Liu et al.* [2011] on the base of comprehensive statistical analysis of electron content determined by radio-occultation technique in COSMIC/FORMOSAT-3 space experiment, for which the quality of ionospheric data is longitudinally independent. It was reported that in the postnoon sector, the average low-latitude NmF2 reached maximal values in two longitudinal ranges: from 90° to 150° and from –180° to –60°, i.e. practically in the same ranges as those for the mean maxVTEC. The good correspondence of the results, obtained by two different ionospheric techniques, is a strong proof of the validity of the techniques used. Hence, our results about longitudinal variations of VTEC obtained from the GIM technique are robust.

Figure 8 shows RGS-related temporal dynamics of the median maxVTEC in comparison with average variations of the solar wind and geomagnetic parameters. Note that the average variation of the solar radiation index F10.7 (not shown) is negligibly weak during RGSs and, thus, it does not affect the average ionospheric variations. In contrast, the solar wind and geomagnetic parameters exhibit prominent average variations during RGSs.

Before RGS (day = -2 and –1), the solar wind velocity and density are low and *B*z ~ 0 that promote a quiet geomagnetic activity and low maxVTEC. The beginning of RGS (day = 0) is caused by a CIR region, in which the solar wind density and negative *B*z reach maximum values that result in intensification of geomagnetic activity and growth of *Kp* and *AE* indices. The increase of solar wind density and velocity prevents a sharp decrease of the *Dst* index. On the second day (day = 1), *B*z persists negative and high while the solar wind velocity increases but the density decreases that results in generation of storm maximum with peak values of the geomagnetic indices. On average, *Kp* and *AE* demonstrate quite well anti-correlation with the *Dst* index.

It is interesting to note that the solar wind velocity peaks on the third day of RGS (day = 2), i.e. on the next day after the RGS maximum. At that time, the southward IMF decreases slightly, the solar wind density approaches to its nominal value of ~5 $cm^{-3}$ and the geomagnetic activity is diminishing. At the same day, the median maxVTEC reaches maximum in practically all the longitudinal ranges. In the range from 150° to 180°, the maximum of median maxVTEC is observed even on the fourth day of RGS (day = 3). On average, the maxVTEC variation correlates better with the solar wind velocity than with the geomagnetic indices. The minimal values of median maxVTEC are observed right before the onset and in the late recovery phase of RGSs. Such dynamics of maxVTEC allow us to introduce so-called "recurrent ionospheric storms" related rather to CIR/HSS solar wind structures than to RGS alone.

## 2.3. Correlations with solar, heliospheric and geomagnetic parameters

*Lei et al.* [2008c] found high correlation between variations in the thermosphere and solar wind and geomagnetic parameters. It is reasonable to expect a high correlation for the ionospheric variations. As a first step, we calculate correlation of daily maxVTEC in 2004 to 2008 with solar wind–magnetosphere coupling functions such as Newell's merging $V^{4/3}B^{2/3}\sin^{8/3}(\theta_c/2)$ and viscous $D^{1/2}V^2$ functions [*Newell et al.*, 2008] and Kuznetsov's merging $V\sqrt{B-Bz}$ and viscous $DV^{3.5}$ functions [*Kuznetsov et al.*, 1993]. Here *V*, *D*, *B*, *Bz*, and $\theta_c$ are, respectively, solar wind velocity, density, IMF strength, Z-component and clock angle. We have found poor correlations: 0.17 (partial correlation coefficients, respectively, 0.07 and 0.16) for the Newell's coupling function and 0.25 (0.24 and 0.12, respectively) for the Kuznetsov's coupling function. Hence, the low-latitude ionosphere would be driven by other parameters.

For further analysis, we represent maxVTEC as a linear function *F*:

$$F = a_0 + a_{\cos}\cos\alpha + a_{1F107}F10.7 + a_{2F107}F10.7^2 + a_{Vsw}Vsw + a_D D + a_{Bz}Bz + a_{AE}AE + a_{Kp}Kp + a_{Dst}Dst$$

(1)

Here we take into account a quadratic dependence of VTEC from F10.7, as a much more accurate approximation than a simple linear dependence [e.g. *Liu and Chen*, 2009]. In order to describe a dependence on the ionosphere illumination by sunlight during various seasons, we introduce a subsidiary parameter, so-called annual angle $\alpha$:

$$\alpha = \pi * ((DOY - DOY1)/(DOY2 - DOY1)), \quad (2)$$

where DOY is a day of year, DOY1 and DOY2 are DOYs of June 22 and December 12, respectively. The cosine of the annual angle ($\cos\alpha$) is close to 0 during equinoxes and approaches to 1 (-1) during northern summer (winter) solstice. The annual angle can be considered as an equivalent of the solar zenith angle relative to the geographic equator at the noon meridian.

For the heliospheric parameters including *V*sw, *D*, *Bz*, we use maximum values determined within the 3 hours before the maxVTEC occurrence. This time includes ~1 hour for the solar wind propagation from the upstream monitor plus 2 hours of the time resolution. For the geomagnetic parameters *Kp* and *Dst*, we use, respectively, maximal and minimal values determined within 9 hours. We also use the maximum of the *AE* index within 6 hours before the maxVTEC occurrence. The 9-hour interval includes 2 hours of the time resolution plus ~7 hours, which we assume to be required for the development of the ionospheric response to the magnetospheric perturbations. The 6-hour interval for the *AE* index is based on an assumption that equatorward neutral winds might require up to ~4 hours to propagate from the auroral region to the equator. The free coefficients are calculated for each longitudinal range by a linear regression method applied for the time profiles of maxVTEC and driving parameters (see Eq. 1) during practically the whole time interval from 2004 to 2008. Hence, the statistics include more than 1700 sets of parameter values during both RGSs and quiet days for 12 longitudinal ranges.

Figure 9 shows longitudinal variation of the correlation between maxVTEC and various parameters. We have found that maxVTEC correlates poorly with *Bz* and *D* (not shown). Hence, we exclude these parameters from further consideration. In addition, the geomagnetic indices *Kp* and *Dst* inter-correlate quite well. However, the 1-hour *Dst* index demonstrates higher correlation with maxVTEC than the 3-hour *Kp* index (not shown). Hence, we also exclude the *Kp* index from further consideration. Finally we analyze the following linear combination of 6 parameters:

$$F_6 = a_0 + F_2 + a_{\cos}\cos\alpha + a_{Vsw}Vsw + a_{AE}AE + a_{Dst}Dst \quad (3a)$$

$$F_2 = a_{0F107} + a_{1F107}F10.7 + a_{2F107}F10.7^2 \quad (3b)$$

As one can see in Figure 9a, the correlation of maxVTEC with other parameters varies significantly (within 3 stat-errors) with longitude. Lowest correlations can be seen in the longitudinal ranges from –120° to -60° and from 60° to 120°. Highest correlations of $r$ ~ 0.8 and $r$ ~ 0.75 can be found, respectively, with $F_6$ and $F_2$ in the longitudinal range from –60° to 60°. In the range from 150° to -120° (i.e. from 150° to 180° and from –180° to –120°), the correlation with $F_6$ is also high ($r$ ~ 0.8) while the correlation with $F_2$ is lower ($r$ ~ 0.7). It seems the dynamics of maxVTEC in the longitudinal range from 150° to -90° is controlled not only by solar radiation but other parameters that become important there.

A relatively high correlation of ~0.3 can be found between maxVTEC and maximal *AE* index in the latitudinal range from –150° to –60° (see Figure 9b). The maximal solar wind velocity has quite low correlation with maxVTEC but this correlation exceeds 0.15 in the longitudinal range from 150° to -90°. More prominent increases in correlation can be found for the *Dst* index (Figure 9c). In the range from 150° to -30°, the anti-correlation of negative *Dst* with positive maxVTEC exceeds 0.35. Hence, in the longitudinal range from 120° to -90° the effect of heliospheric and geomagnetic parameters to the ionosphere is more ordered that produces an additional contribution to the correlation with maxVTEC. It is interesting to point out an effect of annual angle (Figure 9c), which exhibits a sharp increase of anti-correlation ($r$ ~ 0.3) in the latitudinal range from 120° to -120°, in comparison with poor anti-correlation ($r$ < 0.2) at other longitudes. We will discuss this effect later.

Figures 10 and 11 show seasonal variations in the correlation. Highest correlations of maxVTEC with $F_6$ (all the parameters) can be found at longitudes from –60° to 60° during summer (June – August), winter (December – February) and autumn (September – November) seasons (see Figure 10a). It seems that the correlation during spring (winter) is lower (higher) than that during other seasons. In Figure 10b one can clearly see a statistically significant decrease of the correlation with $F_2$ (function of $F10.7$) during the spring season. Perhaps, the decrease of the maxVTEC correlation with $F_6$ results from the dependence on solar radiation.

In Figure 11a, one can see that the spring season is also characterized by relatively high anti-correlation between maxVTEC and the annual angle in the range of longitudes from 120° to -90°. A similar tendency but with lower correlation can be found for the autumn season. Hence, the enhancement of anti-correlation between maxVTEC and cos$\alpha$, demonstrated in Figure 9c, is related to the equinoxes. Another interesting feature in Figure 11a is a relatively high correlation between maxVTEC and cos$\alpha$ in the longitudinal range from –90° to 90° during winter. Note that maxVTEC has highest correlations with the solar radiation index $F10.7$ (see Figure 9a) in almost the same range (from –60° to 60°).

It is important to point out that higher winter (December to February) correlation is also revealed for heliospheric and geomagnetic parameters (see Figure 11). In general, the longitudinal variations of the correlations during different seasons have similar patterns as shown in Figure 9. Namely, the correlations increase in the longitudinal range from 120° to –90°. The highest correlations are found during the period from December to February such that the correlation approaches and exceeds 0.5 and 0.3 for geomagnetic indices and for $V$sw, respectively. Hence, the ionospheric storms correlate better with RGSs around winter solstice.

## 3. Height profiles of disturbed electron content

A comprehensive statistical analysis of the ionospheric response to RGSs was performed on the base of COSMIC/FORMOSAT-3 space-borne data by *Ram et al.* [2010]. They dealt with zonal average electron density and did not distinguish between different phases of RGS related, respectively, to CIR and HSS. In order to study physical mechanisms of generation of positive ionospheric storms at different longitudes during RGSs, however, we consider vertical profiles of the electron density (EC) measured by ionosondes and retrieved from COSMIC/FORMOSAT-3 observations during the HSS-related maximum of RGS and ionospheric storm on 25 March 2008.

We have used the advanced Digital Ionosonde (Digisonde-4D or DPS-4D) database obtained from Kwajalein and from Jicamarca stations. The height profiles of EC were 'True Height Profiles' which are obtained by the ARTIST-5 (Automatic Real-Time Ionogram Scaler with True Height) software. ARTIST-5 software calculates automatically the topside profile based on the bottomside profile of the ionogram by using appropriate extrapolation techniques, so that full profiles of EC can be obtained unambiguously. Further details on ARTIST-5 algorithms can be obtained from *Galkin et al.* [2008].

The COSMIC/FORMOSAT-3 constellation of six low-orbit satellites produces a sounding of the ionosphere using the radio occultation (RO) technique, which makes use of radio signals transmitted by the GPS satellites [*Hajj et al.*, 2000]. Usually over 2500 soundings per day provide EC height profiles at altitudes from the Earth surface to the orbital altitude at ~800 km over ocean and land. A 3-D EC distribution can thus be deduced through relaxation using red-black smoothing on numerous EC height profiles [see *Tsai et al.*, 2006]. This 3-D EC image is used as an initial guess to start the iterative Multiplicative Algebraic Reconstruction Technique (MART) algorithm, and 3-D tomography of the EC is then produced with a time step of 2 hours around whole globe with spatial grid of 5° in longitude, 1° in latitude, and 5 km in height. Such tomographic investigations give valuable information in a wide region, without use of models.

Figure 12 shows the height profile of EC in the Pacific region at ~ 2 UT during the quiet day of 25 March 2008 and in the maximum of RGS-related ionospheric storm on 28 March 2008. The height profiles of EC were measured by an ionosonde located at Kwajalein (9°N, 167°E). As one can clearly see in Figure 12a, the NmF2 increased significantly (more than 50%), the F2 layer is thickening and the hmF2 elevated up to ~ 50 km from the height of ~250 km to ~300 km. From COSMIC/FORMOSAT-3 RO data, we reconstructed a meridional cut of EC in the range from 165° to 170° corresponding to ~13 LT (see Figure 12b). In this cut, the maximum of F2 layer was located in the southern crest of EIA. During the storm maximum, a total EC (TEC), integrated through all heights at latitude of the maximum (lat ~ 5°), increased up to ~18 TECU in comparison with ~13 TECU during the quiet day, i.e. by ~50%. In addition, the hmF2 in the southern crest increased by a few tens of km and the thickness of F2 layer also increased.

Figure 13 shows an example of RGS-related dynamics of the EC height profiles in the South American region. According to ionosonde measurements at Jicamarca (12°S, 77°W), the NmF2 in the postnoon sector (22UT and 17LT) increased significantly and elevated from ~300 km height on the quiet day of 25 March 2008 to ~350 km during the storm on 28 March 2008 (see Figure 13a). Note that hmF2 in the American region is ~50 km higher than that in the Pacific region. The reconstruction of EC meridional cut from COSMIC/FORMOSAT-3 RO data in the range from -80° to -75° (Figure 13b) showed that during the storm, the NmF2 moved from the quiet-day location in the northern crest to the storm-time southern crest. The storm-time F2 layer was much thicker than that at the quiet day. The TEC increased by ~50% in the whole EIA region from quiet values of 15 – 18 TECU to storm-time values of 20 - 24 TECU. At the latitude of maximum (lat ~ -14°), the TEC increased from ~15 TECU to ~22 TECU and the hmF2 elevated from ~290 km to ~310 km. Note that in the northern crest, the hmF2 increased from ~330 km to ~340 km.

Considering Figures 12 and 13, we can also find that the topology of EIA is not changed very much in the storm maximum. In particular, the latitude of the crests remains almost constant. Using results by *Balan and Bailey* [1995], we estimated that ~50 km increase in hmF2 should result in a poleward displacement of the crest regions by about a few degrees, which was hard to distinguish using 1° × 5° angular resolution. Hence, in the maximum of RGS-related ionospheric storm on 28 March 2008, the EIA is intensified: the F2 layer is thickening, the NmF2 increases by ~50% and the hmF2 elevates by a few tens of kilometers.

## 4. Discussion

From analysis of global ionospheric maps during declining phase of the 23-rd solar cycle (2004 – 2008), we have found that recurrent geomagnetic storms result in positive ionospheric storms at low-latitudes on the dayside. The storms are revealed as prominent enhancements of VTEC in the EIA region. We have also found that the amplitude of positive ionospheric storms varies substantially with longitude. The longitudinal variations in the RGS-driven ionosphere are difficult to explain by the effects typical for CME-driven storms such as an effect of "storm start time" [*Fuller-Rowell et al.*, 1994]. As one can see in Figures 5, the RGSs onsets are practically equally probable at different UT. Moreover, RGSs last for several days and largest enhancements of VTEC occur ~2 days after the onset and ~1 day after the storm maximum (see Figure 8) such that the ionosphere losses information not only about the onset but also about the main phase of a storm. In contrast to CME-driven storms, an RGS is generated by two geoeffective solar wind structures: CIR and HSS. The onset and main phase of RGS are related to CIR. The prolonged maximum and recovery phase are related to HSS.

In the beginning of RGS, VTEC in the crest regions of EIA does not exhibit any prominent enhancement and even has a tendency to be suppressed (see Figure 3, 4, 7 and 8). Such behavior can be caused by an increase of recombination rate. *Verkhoglyadova et al.* [2011] reported an excess of thermospheric heating at low to high latitudes during CIR intervals. This excess was interpreted as additional energy input from geomagnetic activity heating. The heating causes a global increase in neutral density, which exceeds 40% at 400 km altitude in the low-latitude dayside ionosphere [*Lei et al.*, 2011; *Liu et al.*, 2012b]. The increase of neutral density does not affect the vertical integral of TEC. However, the increase of temperature results in an increase of the recombination rate and, thus, in a decrease of the ionization.

On the other hand, a continuous increase of $\Sigma O/N_2$ at low latitudes during RGSs was reported [*Crowley et al.*, 2008; *Liu et al.*, 2012b]. The change of chemical composition of neutral species is attributed to dawnward vertical winds, which carry atomic-oxygen-rich air to lower latitudes that result in enhancements in F-region electron densities. The vertical winds are primarily driven by meridional neutral winds blowing toward equator from the hot auroral regions where Joule and particle heating is produced by continuous geomagnetic activity.

Hence in the beginning of recurrent storms, the VTEC dynamics are affected by two opposite factors: atmospheric heating, which increases the recombination rate, and downward vertical winds enriched by atomic oxygen, which increase the electron density. In addition, dayside VTEC enhancements, observed sometimes during CIR intervals, can be produced by strong PPEF of interplanetary origin and/or by the enhancements of solar radiation flux *F*10.7 (see Figures 3 and 4).

The CIR interval is followed by HSS. During HSS, VTEC increases substantially in the maximum of RGSs and peaks during the recovery phase of RGSs on the third day after the onset (see Figure 8). From analysis of EC vertical profiles derived from two independent methods of ionosondes and COSMIC/FORMOSAT-3 RO technique, we have found that the maximum of RGS-related positive ionospheric storm is characterized by ~50% increase in VTEC, elevations of hmF2 by a few tens of kilometers and thickening of the F2 layer, while the latitude of crests remains practically unchangeable. This pattern is very close to that revealed by *Ram et al.* [2010] from COSMIC/FORMOSAT-3 RO measurements of the ionospheric disturbances during RGSs. It was suggested that the observed changes in EC could be related to enhancements in the thermospheric temperature and neutral composition $\Sigma O/N_2$. Note that *Ram et al.* [2010] did not distinguish between CIR and HSS intervals.

In the low-latitude thermosphere, HSS intervals are accompanied by a decrease of the density and temperature [*Verkhoglyadova et al.*, 2011, *Liu et al.* 2012b]. Hence, higher peak and scale heights of the F2 layer in the maximum of the ionospheric storm cannot be explained by the thermospheric factors. However, an excess of the ratio $\Sigma O/N_2$ at low latitudes still persists [*Crowley et al.*, 2008; *Liu et al.* 2012b]. The excess of $\Sigma O/N_2$ results from continuous heating at middle and high latitudes produced by geomagnetic activity [*Sojka et al.*, 2009; *Verkhoglyadova et al.*, 2011], which is generated by high-amplitude IMF variations inherent to HSS. The high ratio $\Sigma O/N_2$ can explain an increase of VTEC but it is unable to explain the changes of peak and scale heights.

The elevation of nmF2 and thickening of the F2 layer as well as increase of VTEC might be also related to intensification of the fountain effect. Note that the effect of DDEF in the noon, postnoon and dusk regions results in suppression or even reversal of the eastward equatorial electrojet and, thus, this mechanism causes a suppression of the fountain effect [*Huang et al.*, 2005; *Huang*, 2012].

The fountain effect on the dayside could be intensified by a prompt penetrating IEF (PPEF) of eastward direction. Studying GIMs with 2-hour resolution does not allow for detecting directly the effect of PPEF. During RGS, the effect is of short duration because of quickly varying IMF *B*z. In Figure 8, one can see that the daily maximum values of negative *B*z remain strong within three days after the RGS onset. Hence, the observed enhancements of 2-hour averaged VTEC might manifest in the operation of eastward PPEF somewhere within the 2-hour interval.

We have found that the maximum of positive ionospheric storm in the crest regions occurs on the day of the maximum solar wind velocity (see Figure 8) on the recovery phase of RGS. The long-lasting recovery phase of RGSs is produced by HILDCAAs, which in turn are related to HSSs characterized by large variations of IEF. Ultra-low-frequency fluctuations of IEF have a substantial effect on global convection and are an important contributor to the large-scale transfer of solar wind energy to the magnetosphere-ionosphere system [*Lyons et al.*, 2009]. Hence, the fountain effect might be intensified by penetrating electric fields of the magnetospheric origin associated with the enhanced convection during HSSs.

The close relationship between the positive ionospheric storms and recurrent geomagnetic activity is revealed in seasonal variations. Namely, we have shown that in March 2008 (Figure 3), the enhancements of VTEC were much stronger than those in January 2007 (Figure 4) despite the fact that the level of solar radiation was almost the same (*F*10.7 ~ 80). That is a manifestation of a semiannual variation in the ionosphere with maxima in equinoxes. During magnetic quiet times, the ionospheric semiannual variation is originated from variation of circulation pattern controlling the atomic/molecular ratio in the thermosphere [*Rishbeth et al.*, 2000]. During geomagnetic disturbances, it is believed that the semiannual variation in the low-latitude ionosphere is contributed by a semiannual variation of geomagnetic activity caused by a strong coupling of the magnetosphere with the heliospheric driving parameters during equinoxes [*Rishbeth and Mendillo*, 2001; *O'Brien and McPherron*, 2002; *Emery et al.*, 2011].

We have also found a strong longitudinal variation both in the mean VTEC and in the magnitude of storm-time VTEC enhancements in the crest regions (see Figure 7). In our analysis, the ionospheric storms are characterized by daily maxVTEC, which occurs at roughly the same local times (noon and postnoon sectors) each day. Hence, we are implicitly sampling the ionosphere in a constant local time reference frame. In such a constant local time frame, various nonmigrating tidal components (either propagating upwards from the lower atmosphere or generated in-situ in the ionosphere) will alias into stationary planetary wave components [e.g. *Rishbeth and Mendillo*, 2001; *Forbes et al.*, 2006; *Liu et al.*, 2010b]. During equinoxes, the typical longitudinal variability associated with lower atmospheric tides should be wave-3 and wave-4. The tide-related waves are mainly attributed to magnetic quiet periods with $Kp < 3$ and the amplitude of these waves is relatively small [*Kil et al.*, 2012]. *Lin et al.* [2007] reported the maximal amplitude of the wave-4 variation in the postnoon sector to be ~3.5 TECU.

As shown in Figure 7, the longitudinal variations of the median and of the average enhancements of maxVTEC are about 2 and 3 TECU, respectively. In other words, the average longitudinal variations of maxVTEC during RGSs are comparable with the maximal amplitude of the tide-related wave-4. However, it is difficult to find the wave-3 and wave-4 in Figures 7, 9, 10 and 11, which seem to be mostly wave-1 and wave-2. Namely, in Figure 7 one can distinguish a prominent minimum in the longitudinal range from –60° to 60° and one or two maxima at other longitudes. In Figures 9, 10, 11, we can find either a wave-1 with single maximum and minimum or wave-2 with two maxima and minima. This is not entirely unexpected as past studies have found that the wave-3 and wave-4 components in GIM TECs tend to be underestimated compared to those retrieved from satellite observations [*Jee et al.*, 2010; *Chang et al.*, 2013].

Figure 6 shows that the storm-time variations of maxVTEC can often exceed 10 TECU. It means that the effect of lower atmospheric tides alone is insufficient to explain the strong longitudinal variation of the VTEC during RGSs and, thus, other mechanisms should be considered. The lack of wave-3 and wave-4 might be explained by the geomagnetic nature of the VTEC variations. It is also consistent with the heliospheric and geomagnetic variables that we are correlating. Those variables, responsible for the geomagnetic disturbances, cannot capture the wave-3 and wave-4 perturbations produced by lower atmospheric sources.

We have found that during RGSs, the wave-1 and wave-2 are most common and prominent statistically. The positive ionospheric storms have been found to be much stronger in the longitudinal ranges from –150° to –90° and from 90° to 150° than those in the range from –60° to 60° (see Figure 7). We have also found two distinct longitudinal ranges, where parameters of different origin play a major role. Namely, in the range of longitudes from –90° to 90°, maxVTEC correlates better with the solar radiation parameter *F*10.7. In the longitudinal range from 90° to -90°, the correlation with the solar radiation decreases dramatically and the influence of heliospheric and geomagnetic parameters becomes important. We have to point out that this pattern of longitudinal variation does not depend much on the season.

Similar longitudinal variation with a prominent minimum in the longitudinal range from –60° to 60° was found by *Liu et al.* [2011] for the average low-latitude NmF2 derived from COSMIC/FORMOSAT-3 RO measurements. It is important to point out that those space-borne measurements did not suffer from the non-uniform longitudinal coverage. *Liu et al.* [2011] also demonstrated that hmF2 in the American sector is higher by several tens of kilometers than that in the Pacific region. In Figure 13, we have found the same difference. It was suggested that the wave-like longitudinal feature in the mean NmF2 and hmF2 is most likely associated with the ionosphere-atmosphere couplings with sources of lower atmospheric origins, such as the migrating diurnal tides and planetary waves propagating upward to the ionosphere [*Immel et al.*, 2006; *Liu et al.*, 2011]]. However, atmospheric tides cannot explain the maxVTEC variations in relation to the geomagnetic activity, which does not affect the lower atmosphere.

*Kil et al.* [2012] ruled out the possibility that the tidal modulation of the dynamo electric fields in the ionosphere is the source of the wave-1 and wave-2 components of longitudinal variations. They suggested mechanisms of the solar zenith angle and magnetic declination, which controls neutral winds and neutral composition at ionospheric heights [*Rishbeth*, 1998]. The effect of solar zenith angle is originated from a difference in the illumination of a dip equator, which governs the EIA region. The dip equator is defined as a line at which the vector of geomagnetic field at given altitude (say 300 km) is strictly horizontal. Because the geomagnetic dipole is tilted and shifted relative to the axis of Earth's rotation, the dip coordinates do not coincide with the geographic ones and, thus, the ionospheric dynamics suffers from the effects of magnetic dip and declination [e.g. *Challinor and Eccles*, 1971].

In addition, our statistical analysis reveals such annual variations as December to February (Dec-Feb) anomaly and spring asymmetry. The seasonal patterns of the ionospheric variations are explained by chemical and dynamic processes through changes in solar zenith angle, thermospheric composition and global circulations [*e.g. Liu et al.*, 2011]. For the spring asymmetry, we have found a lower correlation of maxVTEC with F10.7 and higher anti-correlation with the annual angle. This asymmetry is most prominent in the longitudinal range from 120° to –90°. It seems that in this range, the solar zenith angle plays a more important role during the equinoxes than during solstices. Because of that the lower correlation with the solar radiation index *F*10.7 is "compensated" by higher anti-correlation with the annual angle.

The Dec-Feb anomaly is characterized by higher correlations of maxVTEC with the annual angle and heliospheric and geomagnetic parameters in December - February. Higher correlations with the annual angle, revealed in the longitude range from –90° to 90°, can be explained in the frame of solar zenith angle effect [*e.g. Kil et al.*, 2012]. For the heliospheric and geomagnetic parameters, the Dec-Feb anomaly consists in highest correlations with maxVTEC in the range of longitudes from ~120° to -90°.

Morphologically, the Dec-Feb anomaly is close to an effect of so-called annual asymmetry revealed for quiet geomagnetic conditions [e.g. *Mendillo et al.*, 2005]. Namely, average noon TEC in December substantially exceeds that in June, almost everywhere around the Globe. At low latitudes, the most prominent asymmetry was reported in the longitudinal range from 120° to –60°, i.e. practically in the same range as that for the Dec-Feb effect in maxVTEC. *Mendillo et al.* [2005] concluded that global changes in the neutral atmosphere (atomic/molecular ratio $O/N_2$) contribute to the ionospheric annual asymmetry. On the other hand, discussing the driven factors of longitudinal variations of the annual asymmetry, *Zeng et al.* [2008] pointed out the effects of solar zenith angle, magnetic field configuration and zonal neutral winds.

During RGSs, the value of $\Sigma O/N_2$ increases at low latitudes for several days [*Crowley et al.*, 2008; *Liu et al.*, 2012b] that should strengthen the Dec-Feb effect. Hence, higher correlation of maxVTEC with the geomagnetic activity around winter solstice can be related to the storm-time enhancement of $\Sigma O/N_2$ at low latitudes. It might be also possible that meridional equatorward neutral winds enforce zonal neutral winds such that the effect of magnetic declination results in additional contribution to the low-latitude electron density enhancements.

It is important to point out that numerous statistical studies of longitudinal variations in the ionosphere did not eliminate the recurrent geomagnetic activity or used a criterion of $Kp < 3$. As one can see in Figure 8, the maximum of positive ionospheric storms corresponds to the average $Kp \sim 3$. Hence, it is important to estimate the contribution of RGSs to the statistics of ionospheric disturbances. For this purpose, we calculate an integral probability of occurrence of solar wind streams with various velocities from 250 to 800 km/s for the time interval from 2004 to 2008 (see Figure 14). Note that the average solar wind speed is ~400 km/s [*Veselovsky et al.*, 2010]. HILDCAAs are related to solar streams with speed >400 km/s [*Tsurutani et al.*, 1995]. In Figures 3 and 4, one can clearly see that positive ionospheric storms persist during whole intervals of HILDCAAs and even 1 to 2 days longer. Hence, we can estimate approximately that the occurrence of positive ionospheric storms corresponds to the occurrence of >400 km/s solar wind streams. In Figure 14, we find that the probability of such occurrence is 0.6. Note that the integral occurrence probability decreases gradually such that the occurrence of solar wind streams with higher speeds of >500 km/s is still a significant portion (~30%) of the whole statistics.

In Figure 14, we also show a scatter plot of 3-hour *K*p index versus solar wind speed *V*sw. On average, *K*p increases with *V*sw. However, the spread is very wide. There are numerous events with small *K*p accompanied by high solar wind speed. We have estimated that 33% (39%) of statistics is characterized by *K*p < 3 (*K*p ≤ 3) under *V*sw > 400 km/s. Such conditions are proper for RGSs and related ionospheric disturbances. As a result, more than >30% of “quiet” ionospheric conditions determined from Kp < 3 can be actually contributed by ionospheric disturbances originated from HSSs.

From the above, we realize that the recurrent ionospheric storms at low latitudes result from a complex of competitive phenomena such as thermospheric heating, change of neutral composition $\Sigma O/N_2$, PPEF, penetrating electric fields of the magnetospheric origin related to HILDCAAs, effect of solar zenith angle and probably magnetic declination effect. On the declining phase of solar activity, CIR/HSS result in ionospheric disturbances lasting up to 60% of time or even more. Hence, the recurrent ionospheric storms dominate in the statistics of ionospheric conditions. In this sense, both the equinox maxima in the ionospheric ionization and the Dec-Feb effect of enhanced correlation of TEC with the geomagnetic activity are strongly related to ionospheric disturbances during RGSs.

## 5. Conclusions

Analysis of case events and statistics of positive ionospheric storms in the region of dayside equatorial ionization anomaly during recurrent geomagnetic storms on the declining phase of the 23rd solar cycle (years from 2004 to 2008) reveals the following:

1. During the storm onset related to a co-rotating interaction region, the total electron content increases slightly or even decreases. The suppression might be caused by a strong heating of the low-latitude thermosphere.

2. Positive ionospheric storms appear as substantial enhancements of the total electron content within a few days (1 – 4 days) after the storm onset during high-intensity long-duration continuous auroral activity produced by high-speed solar wind streams. On average, the maximum of positive ionospheric storms occurs 3 days after the onset and on the recovery phase of recurrent magnetic storms, at the day of maximum solar wind velocity.

3. In the maximum of ionospheric storm occurred during Whole Heliospheric Interval and related to HSS, we find that at substantially different longitudes, the F2 layer is thickening, NmF2 increases by ~50% and hmF2 elevates by a few tens of kilometers. We speculate that this might result from the increase of $\Sigma O/N_2$ and penetrating electric fields of interplanetary and magnetospheric origin.

4. Statistically, the response of positive ionospheric storms to solar, heliospheric and geomagnetic drivers has a prominent longitudinal asymmetry. In the longitudinal range from –90° to 90°, the solar illumination plays a major role. In the range of longitudes from 90° to 180° and from –180° to –120°, the influence of heliospheric and geomagnetic drivers becomes significant.

5. Months from December to February are characterized by the highest correlations of the ionospheric ionization enhancements with the heliospheric and geomagnetic drivers.

**Acknowledgements** The authors thank Kyoto World Data Center for Geomagnetism (http://swdcwww.kugi.kyoto-u.ac.jp/index.html) for providing the *Dst*, *Kp* and *AE* geomagnetic indices. The Global Ionosphere Maps/VTEC Data are produced by European Data Center (http://www.aiub.unibe.ch/ionosphere/). The ACE solar wind data were provided by N. Ness and D.J. McComas through the CDAWeb website. We thank Dr. N. Balan for very useful discussion and valuable recommendations. This work was supported by grants NSC-100-2119-M-008 -019- from the National Science Council of Taiwan and by Ministry of Education under the Aim for Top University program at National Central University of Taiwan.

Table 1. Number of low- to mid-latitude IGS receivers in different longitudinal ranges

| -180° | -150° | -120° | -90° | -60° | -30° | 0° | 30° | 60° | 90° | 120° | 150° |
|---|---|---|---|---|---|---|---|---|---|---|---|
| -150° | -120° | -90° | -60° | -30° | 0° | 30° | 60° | 90° | 120° | 150° | 180° |
| 6 | 4 | 8 | 10 | 6 | 6 | 10 | 11 | 9 | 10 | 11 | 5 |

## Figure Captions

Figure 1. Solar-cycle variations of solar radio flux F10.7 (top panel), solar wind speed (middle panel) and daily minimum Dst index (bottom panel). The interval used for analysis from 2004 to 2008 (indicated by red) corresponds to declining phase and solar minimum accompanied by co-rotating high-speed solar wind streams and recurrent geomagnetic storms (RGSs) of moderate and low intensity (Dstmin > -70 nT as indicated by the dashed line at the bottom panel).

Figure 2. Global ionospheric maps (GIMs) of vertical total electron content (VTEC) constructed at 22 UT (left) and 8 UT (right) during (a) quiet day on 25 March 2008, (b) disturbed days on 27 (left) and 29 (right) March 2008, and (c) residual between the disturbed and quiet days. March 27 and 29 correspond, respectively, to maximum and recovery phase of the storm (see Figure 3). The dip equator is shown by the white curve. Vertical black dashed lines indicate local noon. Strong positive ionospheric storms occur at low latitudes in the postnoon and evening sectors in Indochina, Pacific, and American regions (at longitudes from 60° to –60°).

Figure 3. 27-day interval of solar, heliospheric, geomagnetic and ionospheric variations from 20 March to 15 April 2008 (from top to bottom): solar radio flux F10.7, solar wind velocity, density, IMF Bz component in GSM, geomagnetic indices AE, Dst, daily maximum VTEC (solid curve) and variations of VTEC (colored 2D histogram) in longitudinal ranges from -120° to -90°, from 90° to 120° and from –30° to 0°. The black stars indicate maxima of VTEC (maxVTEC). The red stars indicate the corresponding magnitude of external parameters (see details in the text). The red vertical arrows indicate the onset of RGSs. The blue circles depict suppressions of maxVTEC in the beginning of RGSs. Peak values of maxVTEC are indicated by the red circles. maxVTEC increases

both during enhancements of the solar radiation (F10.7) and during RGSs. In different latitudinal ranges, maxVTEC increases in different manner both temporally and spatially.

Figure 4. The same as in Figure 3 but for the interval from 1 January to 3 February 2007 and longitudinal ranges for maxVTEC from -150° to -120°, form 120° to 150° and from 0° to 30°. Variations of maxVTEC exhibit similar spatial and temporal patterns.

Figure 5. Universal time distribution for the onsets of 185 recurrent geomagnetic storms occurred from 2004 to 2008. The RGS onsets are practically equally probable at different UT.

Figure 6. Superposed epoch analysis of *Dst* (top panel) and maxVTEC variations at longitudes from 0° to 30° (middle panel) and from 120° to 150° (bottom panel) during 185 RGSs occurred from 2004 to 2008. Variations during different storms are shown by thin curves. The daily medians are indicated by thick red curves. On the top panel, the white dashed and red solid curves depict the median of hourly averaged and of daily minimum *Dst*, respectively. The onset of RMS corresponds to day = 0. The amplitude of median maxVTEC variations in the longitudinal range from 120° to 150° is about 30% higher than that in the range from 0° to 30°.

Figure 7. Longitudinal variations of average maxVTEC characteristics during RGSs occurred from 2004 to 2008. Upper panel shows the amplitudes of median maxVTEC variations (solid histogram, left axis) and variation of median maxVTEC during RGS onsets (day = 0) (dashed histogram, right axis). Lower panel shows the mean maxVTEC (solid histogram, left axis) and average enhancements of maxVTEC (dashed histogram, right axis). Bold numbers indicate the amount of receivers in each longitudinal range. The longitudinal range from -60° to 60° is characterized by lowest variability of maxVTEC.

Figure 8. Average variations of various parameters during 185 RGSs occurred from 2004 to 2008 (from top to bottom): maximal solar wind velocity (solid curve, left axis) and density (dashed curve, right axis); minimal Dst (solid curve, left axis) and maximal Kp (dashed curve, right axis); maximal AE (solid curve, left axis) and minimal Bz (dashed curve, right axis); median maxVTEC in various longitudinal ranges (solid and dashed curves correspond to eastern and western longitudes, respectively). The onset of RMS corresponds to the day = 0 (vertical dashed line). Geomagnetic indices peak one day after the onset. Maximum of the solar wind velocity and maxVTEC occur 2 days after the onset (indicated by vertical solid line).

Figure 9. Correlation of maxVTEC with various parameters: (a) all set of parameters and a function of F10.7, (b) AE index and solar wind velocity, (c) Dst index and cosine of annual angle (cos).

Figure 10. Seasonal variation of the correlation between the maxVTEC and (a) all set of parameters and (b) a function of F10.7. maxVTEC has a weaker correlation with F10.7 during spring equinox at all longitudes.

Figure 11. Seasonal variation of the correlation between the maxVTEC and (a) cosine of annual angle, (b) Dst index, (c) solar wind velocity, and (d) AE index and. The cosine of annual angle shows a strong anti-correlation around spring equinox. High correlations with the solar wind and geomagnetic parameters are found during winter (Dec – Feb), especially in the Pacific and South America regions.

Figure 12. Height profile of electron content (EC) measured in the Pacific region (a) over Kwajalein (9°N, 167°E) and (b) reconstructed from COSMIC/FORMOSAT-3 radio-occultation tomography at longitude 167°E during a quiet day on 25 March 2008 (right panels) and in the maximum of RGS-related positive ionospheric storm on 28 March 2008 (left panels). During the storm, the EC in the Pacific region increases substantially and the maximum of F-layer elevates up to ~ 50 km from the height of ~250 km to ~300 km.

Figure 13. The same as in Figure 12, but over Jicamarka (12°S, 77°W) and at longitude 80°W. During the storm, the EC in the South America region increases substantially and the maximum of F-layer elevates up to ~ 50 km from the height of ~300 km to ~350 km.

Figure 14. Integral probabilities: (solid curves, left axis) of occurrence of solar wind streams with velocities higher than given and (green isolines with numbers, right axis) of *Kp* smaller than given at various velocities in 2004 – 2008. Dots depict the scatter plot of *Kp* versus speed. High-speed solar wind streams with velocity >400 km/s (depicted by red curve) occur in 60% of statistics. More than 30% of statistics is characterized by small Kp < 3 under high-speed solar wind streams that is proper for RGSs.

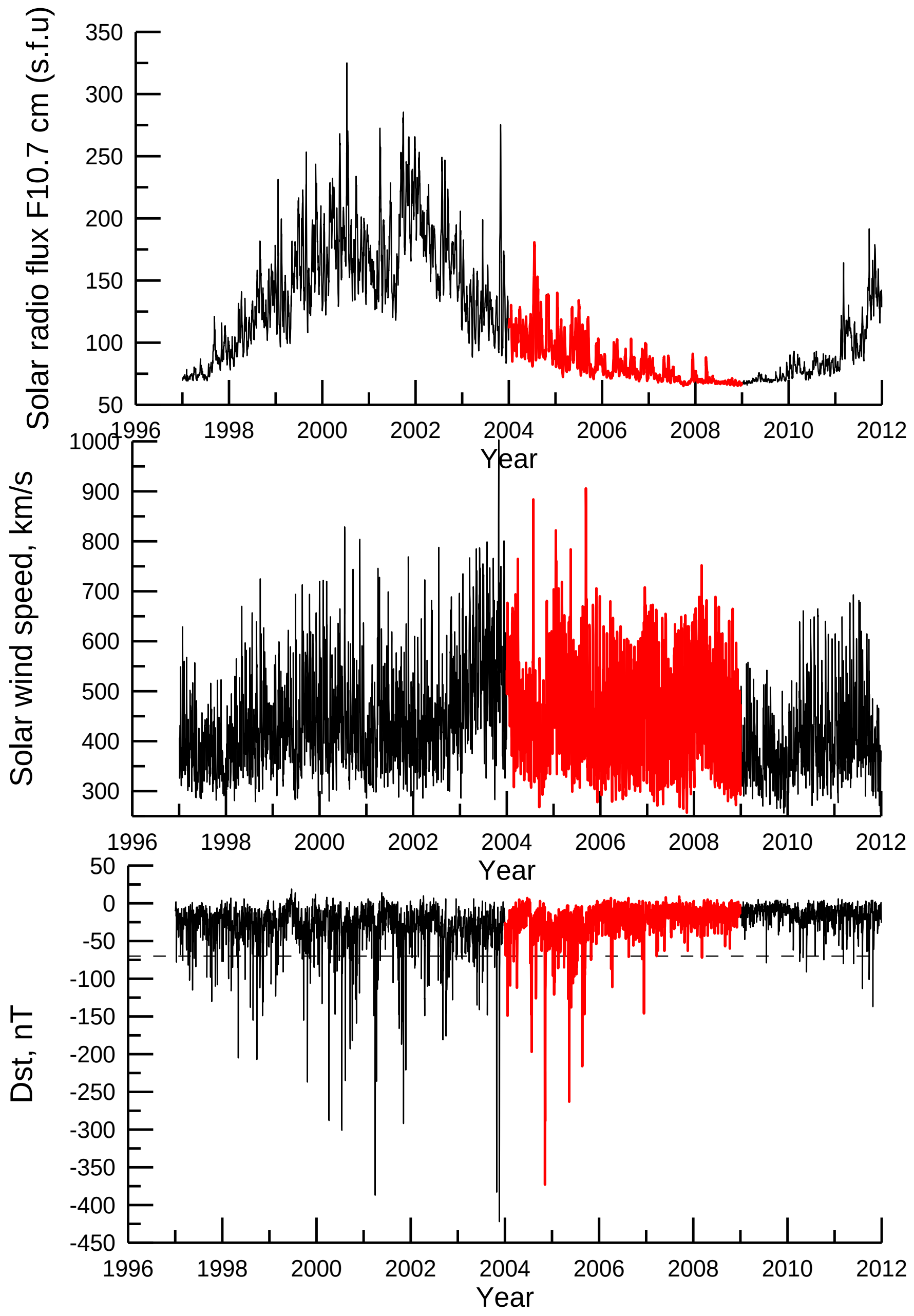

Figure 1. Solar-cycle variations of solar radio flux F10.7 (top panel), solar wind speed (middle panel) and daily minimum Dst index (bottom panel). The interval used for analysis from 2004 to 2008 (indicated by red) corresponds to declining phase and solar minimum accompanied by co-rotating high-speed solar wind streams and recurrent geomagnetic storms (RGSs) of moderate and low intensity (Dstmin > -70 nT as indicated by the dashed line at the bottom panel).

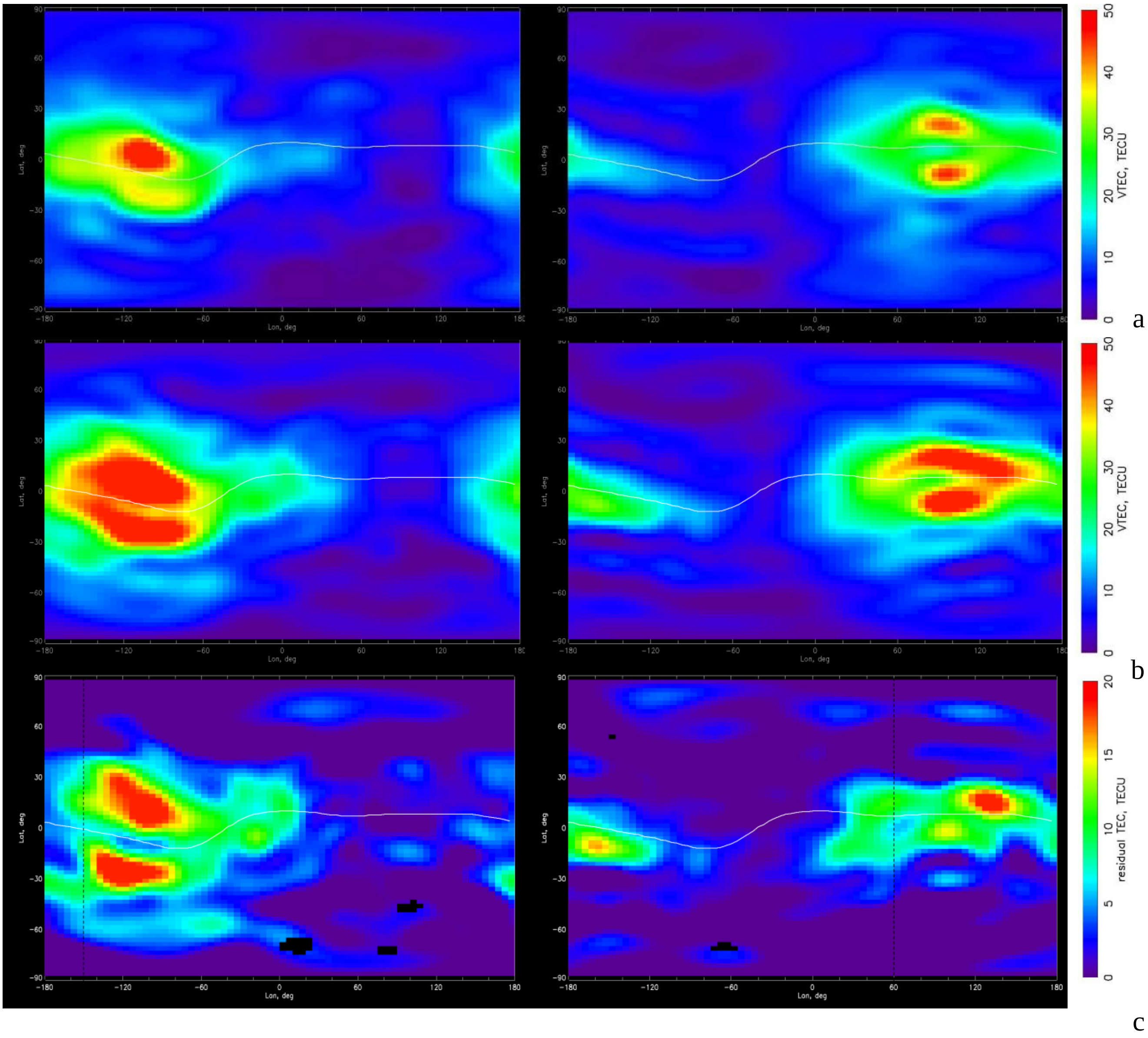

Figure 2. Global ionospheric maps (GIMs) of vertical total electron content (VTEC) constructed at 22 UT (left) and 8 UT (right) during (a) quiet day on 25 March 2008, (b) disturbed days on 27 (left) and 29 (right) March 2008, and (c) residual between the disturbed and quiet days. March 27 and 29 correspond, respectively, to maximum and recovery phase of the storm (see Figure 3). The dip equator is shown by the white curve. Vertical black dashed lines indicate local noon. Strong positive ionospheric storms occur at low latitudes in the postnoon and evening sectors in Indochina, Pacific, and American regions (at longitudes from 60° to –60°).

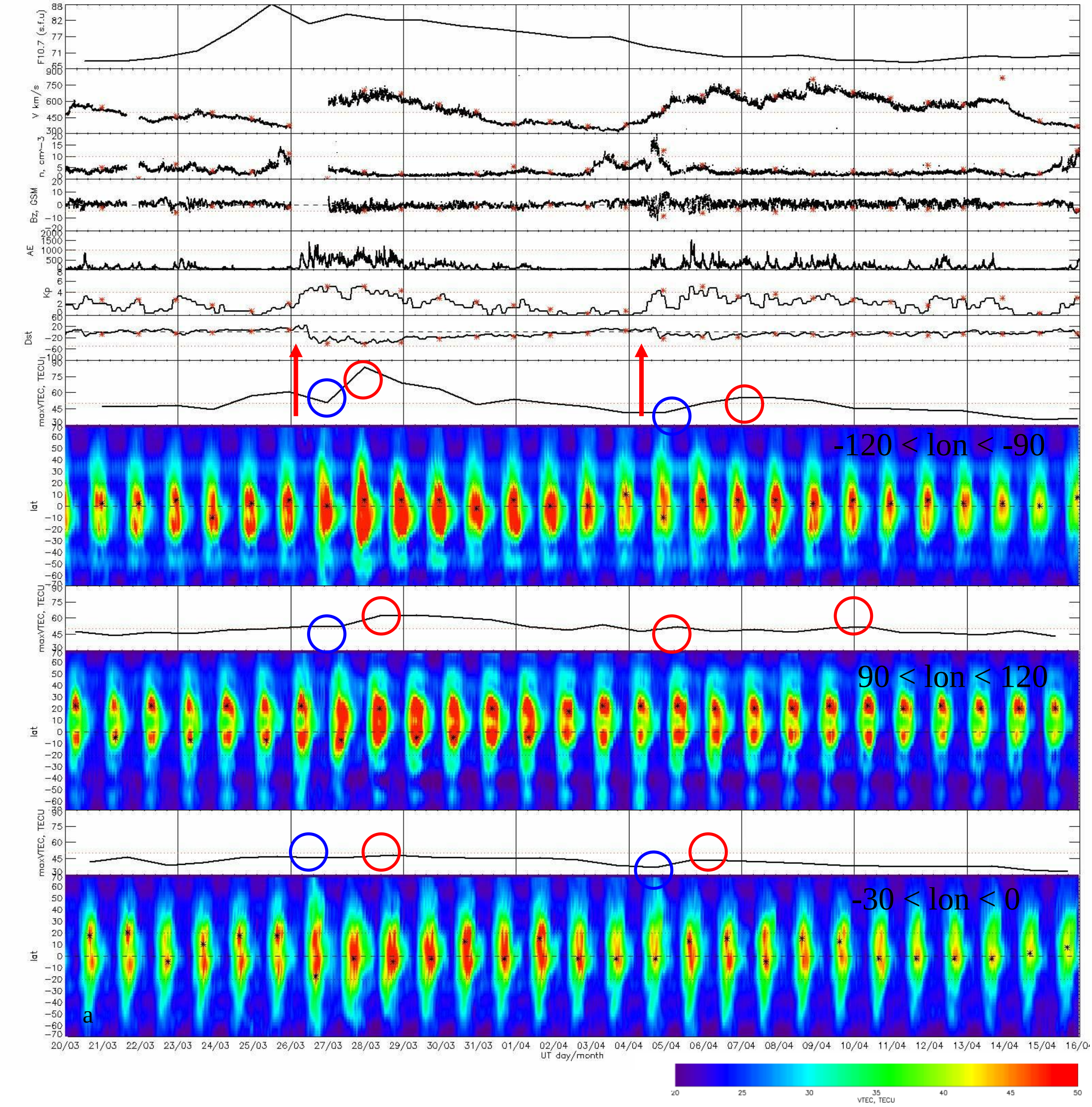

Figure 3. 27-day interval of solar, heliospheric, geomagnetic and ionospheric variations from 20 March to 15 April 2008 (from top to bottom): solar radio flux F10.7, solar wind velocity, density, IMF Bz component in GSM, geomagnetic indices AE, Dst, daily maximum VTEC (solid curve) and variations of VTEC (colored 2D histogram) in longitudinal ranges from -120° to -90°, from 90° to 120° and from –30° to 0°. The black stars indicate maxima of VTEC (maxVTEC). The red stars indicate the corresponding magnitude of external parameters (see details in the text). The red vertical arrows indicate the onset of RGSs. The blue circles depict suppressions of maxVTEC in the beginning of RGSs. Peak values of maxVTEC are indicated by the red circles. maxVTEC increases both during enhancements of the solar radiation (F10.7) and during RGSs. In different latitudinal ranges, maxVTEC increases in different manner both temporally and spatially.

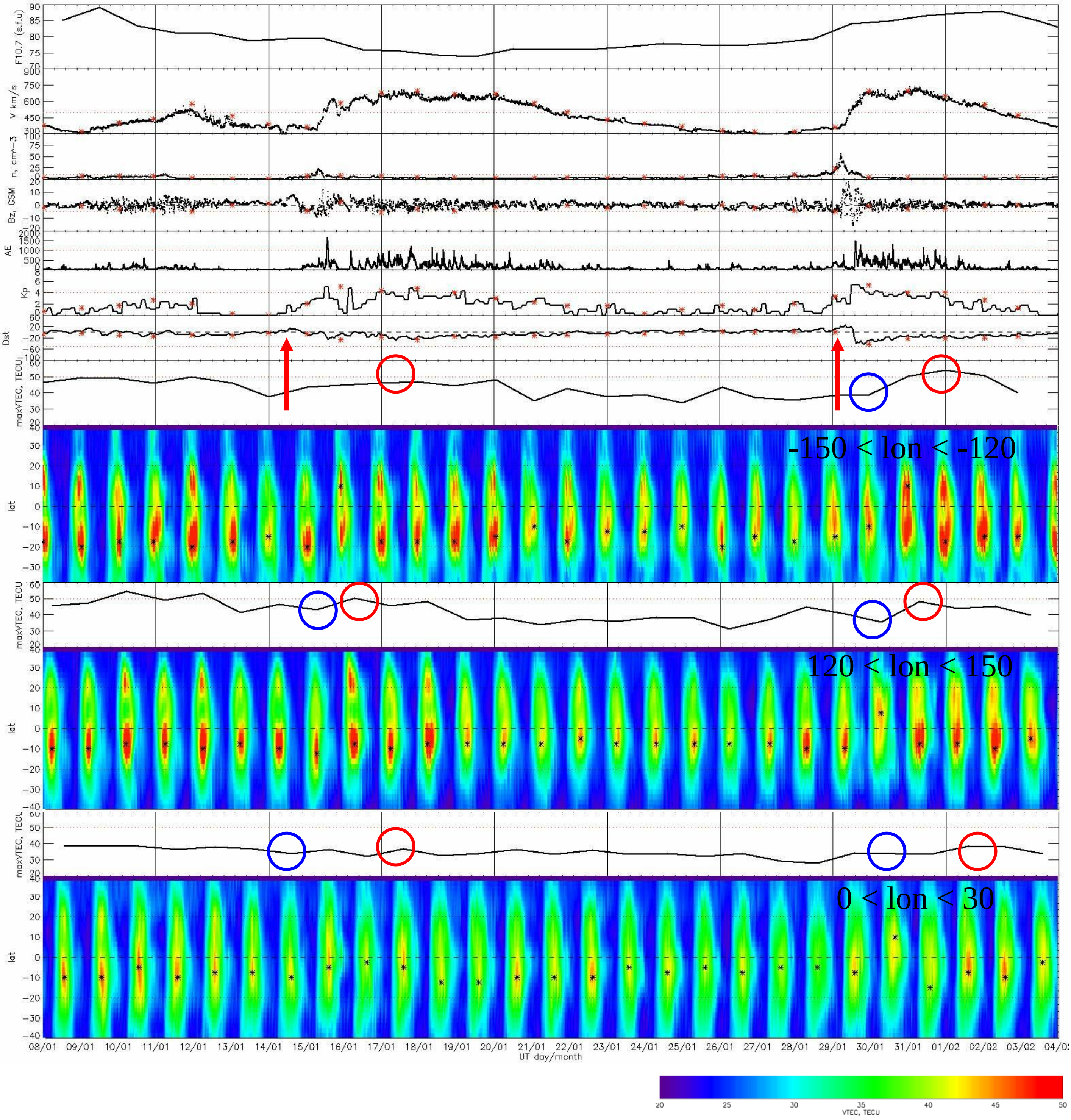

Figure 4. The same as in Figure 3 but for the interval from 1 January to 3 February 2007 and longitudinal ranges for maxVTEC from -150° to -120°, form 120° to 150° and from 0° to 30°. Variations of maxVTEC exhibit similar spatial and temporal patterns.

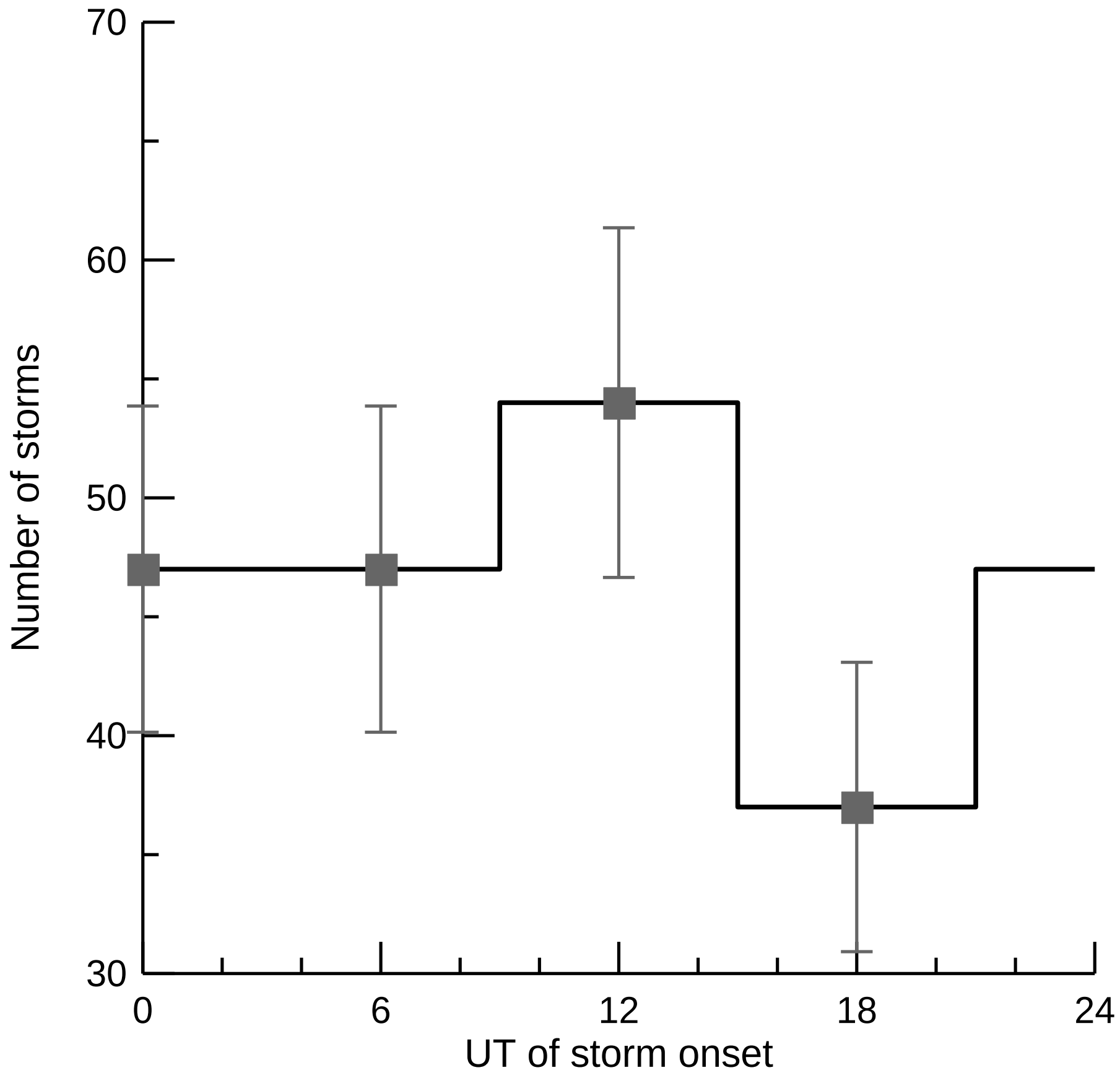

Figure 5. Universal time distribution for the onsets of 185 recurrent geomagnetic storms occurred from 2004 to 2008. The RGS onsets are practically equally probable at different UT.

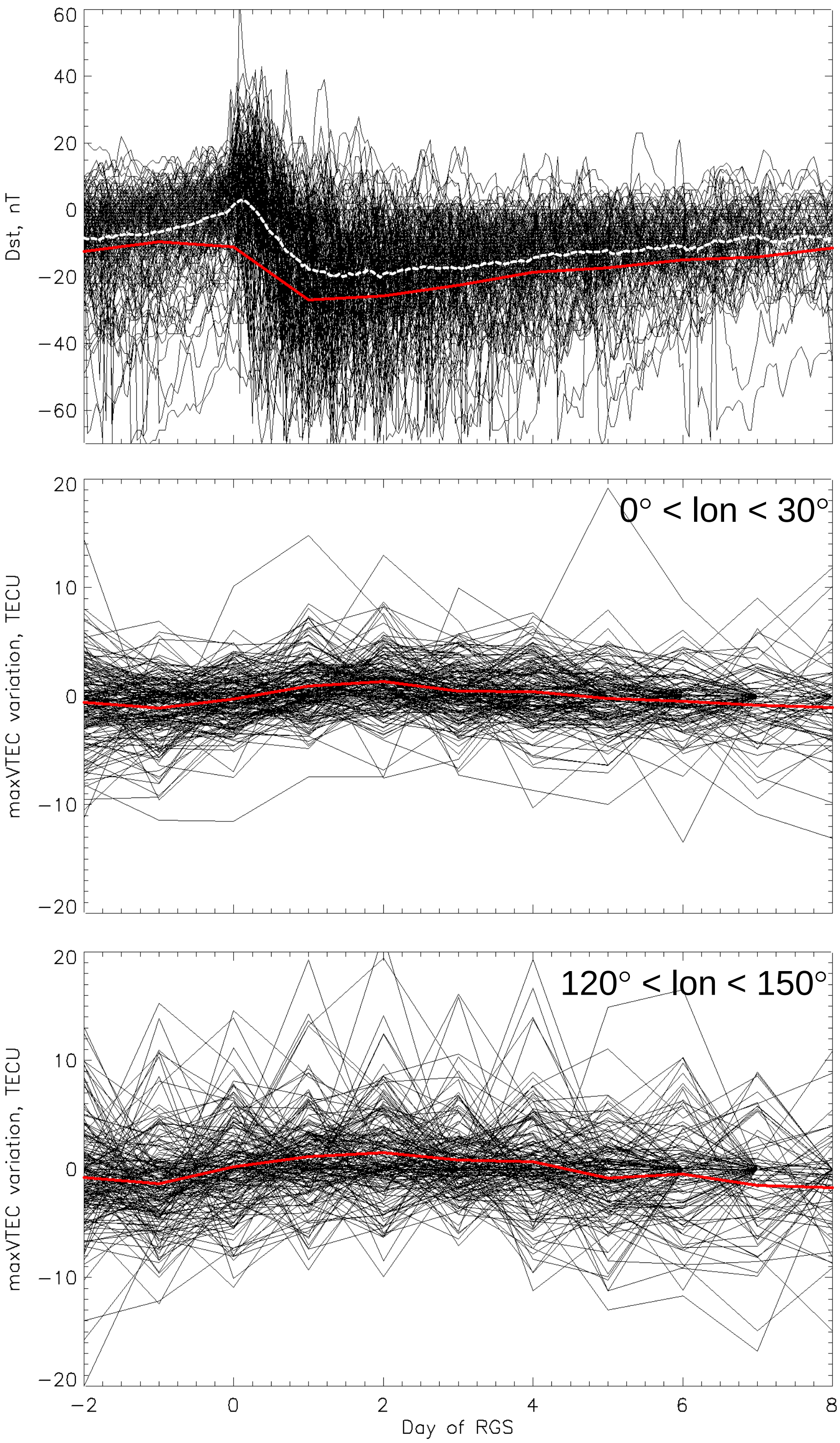

Figure 6. Superposed epoch analysis of *Dst* (top panel) and maxVTEC variations at longitudes from 0° to 30° (middle panel) and from 120° to 150° (bottom panel) during 185 RGSs occurred from 2004 to 2008. Variations during different storms are shown by thin curves. The daily medians are indicated by thick red curves. On the top panel, the white dashed and red solid curves depict the median of hourly averaged and of daily minimum *Dst*, respectively. The onset of RMS corresponds to day = 0. The amplitude of median maxVTEC variations in the longitudinal range from 120° to 150° is about 30% higher than that in the range from 0° to 30°.

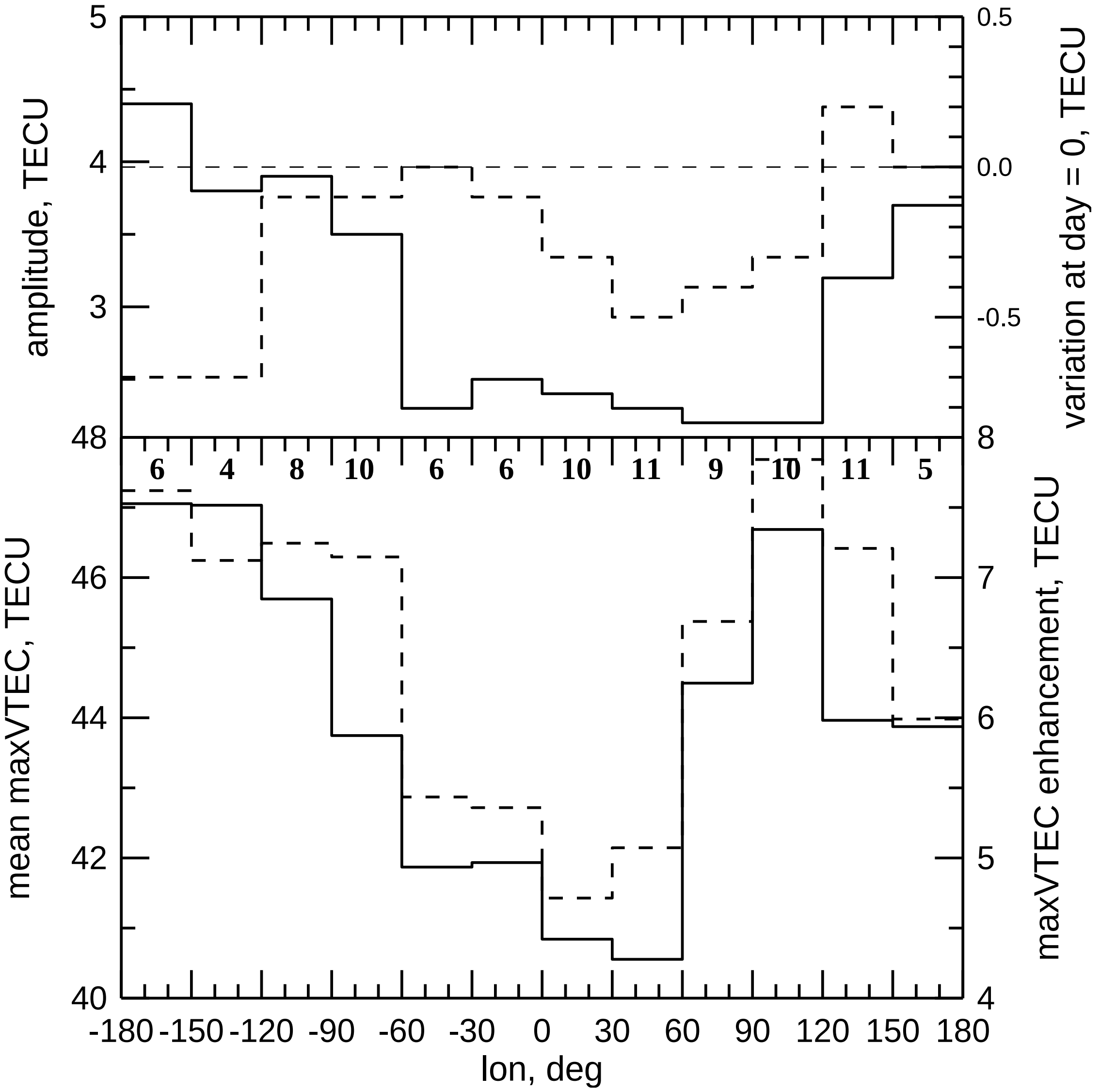

Figure 7. Longitudinal variations of average maxVTEC characteristics during RGSs occurred from 2004 to 2008. Upper panel shows the amplitudes of median maxVTEC variations (solid histogram, left axis) and variation of median maxVTEC during RGS onsets (day = 0) (dashed histogram, right axis). Lower panel shows the mean maxVTEC (solid histogram, left axis) and average enhancements of maxVTEC (dashed histogram, right axis). Bold numbers indicate the amount of receivers in each longitudinal range. The longitudinal range from -60° to 60° is characterized by lowest variability of maxVTEC.

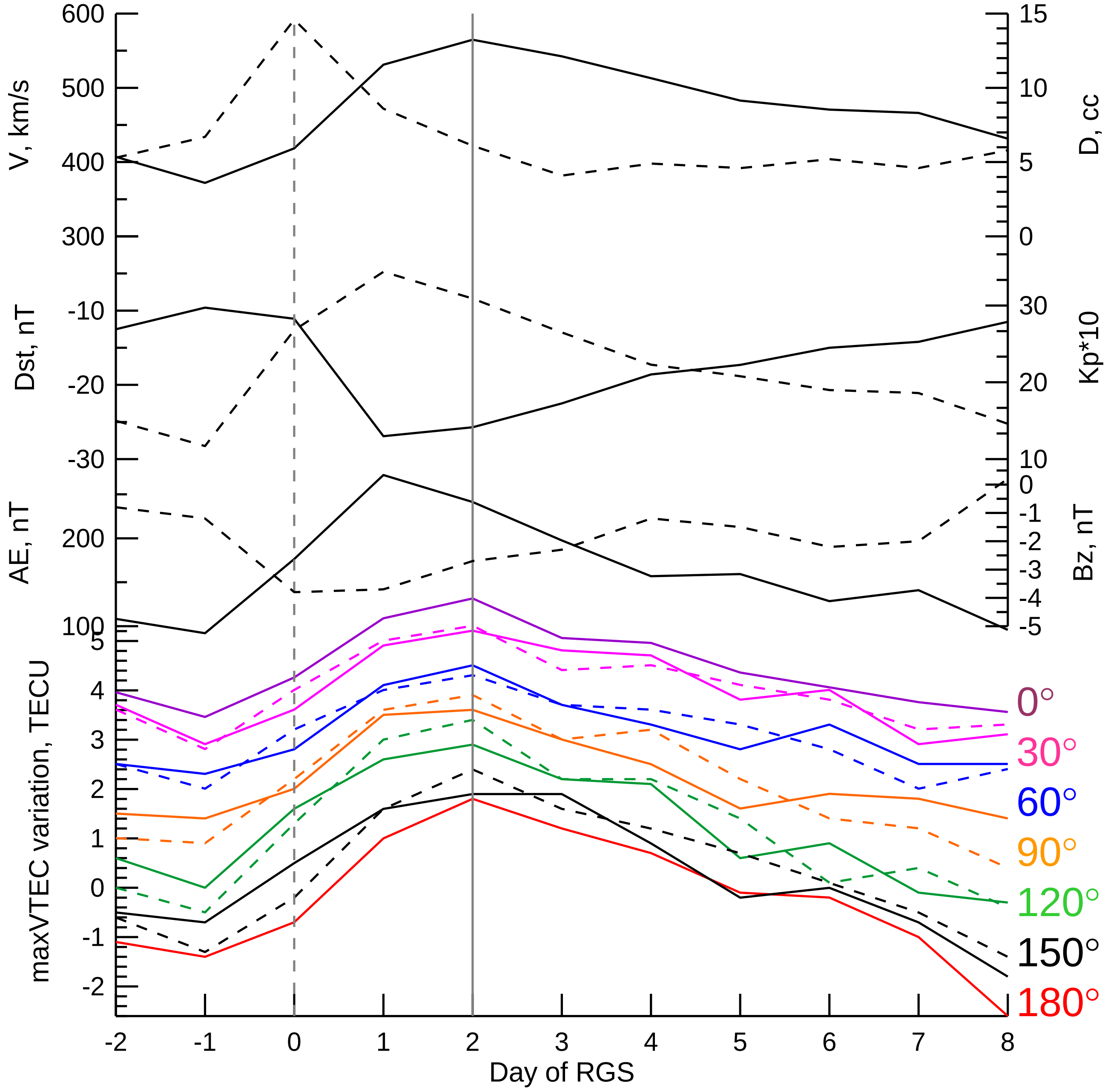

Figure 8. Average variations of various parameters during 185 RGSs occurred from 2004 to 2008 (from top to bottom): maximal solar wind velocity (solid curve, left axis) and density (dashed curve, right axis); minimal Dst (solid curve, left axis) and maximal Kp (dashed curve, right axis); maximal AE (solid curve, left axis) and minimal Bz (dashed curve, right axis); median maxVTEC in various longitudinal ranges (solid and dashed curves correspond to eastern and western longitudes, respectively). The onset of RMS corresponds to the day = 0 (vertical dashed line). Geomagnetic indices peak one day after the onset. Maximum of the solar wind velocity and maxVTEC occur 2 days after the onset (indicated by vertical solid line).

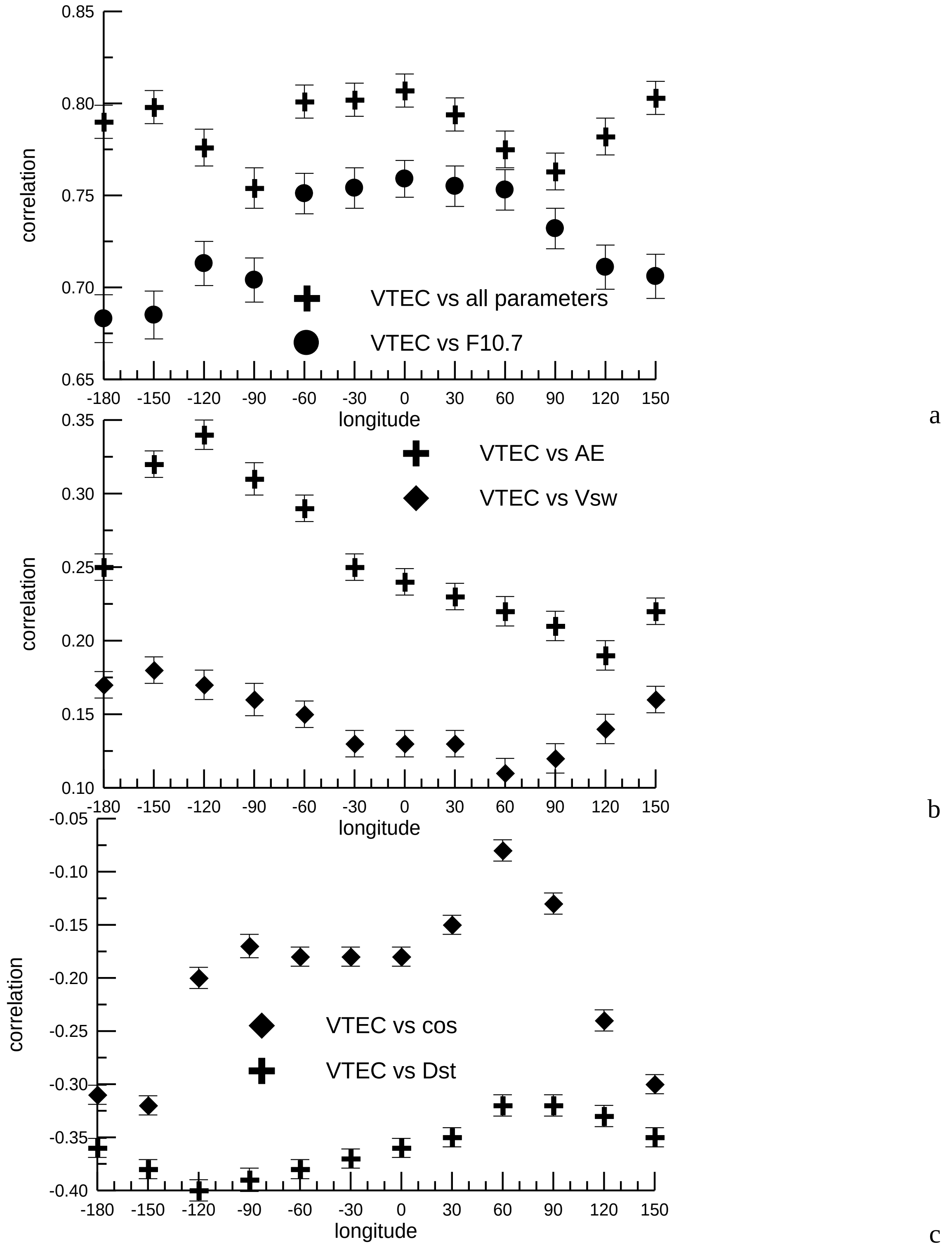

Figure 9. Correlation of maxVTEC with various parameters: (a) all set of parameters and a function of F10.7, (b) AE index and solar wind velocity, (c) Dst index and cosine of annual angle (cos).

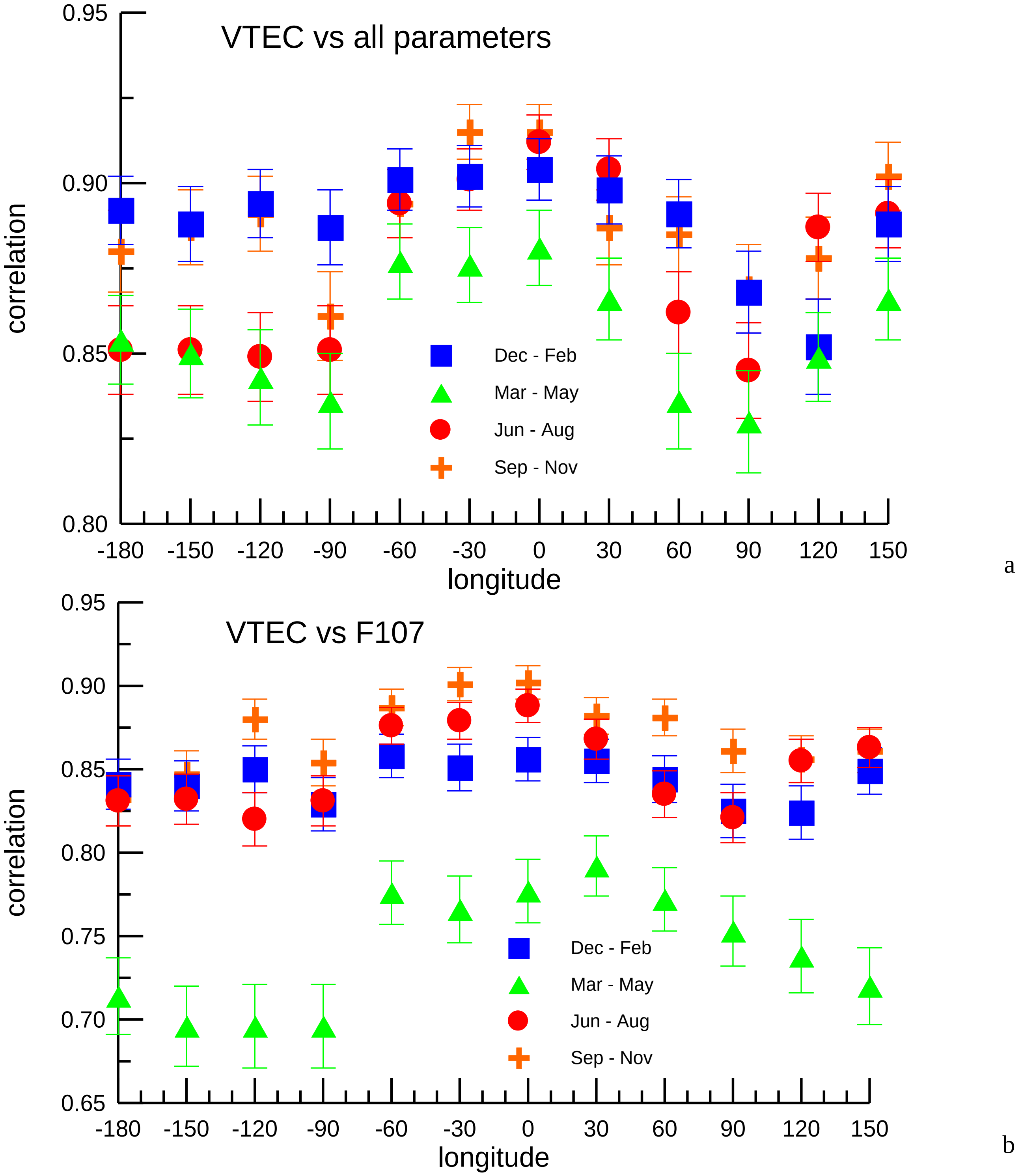

Figure 10. Seasonal variation of the correlation between the maxVTEC and (a) all set of parameters and (b) a function of F10.7. maxVTEC has a weaker correlation with F10.7 during spring equinox at all longitudes.

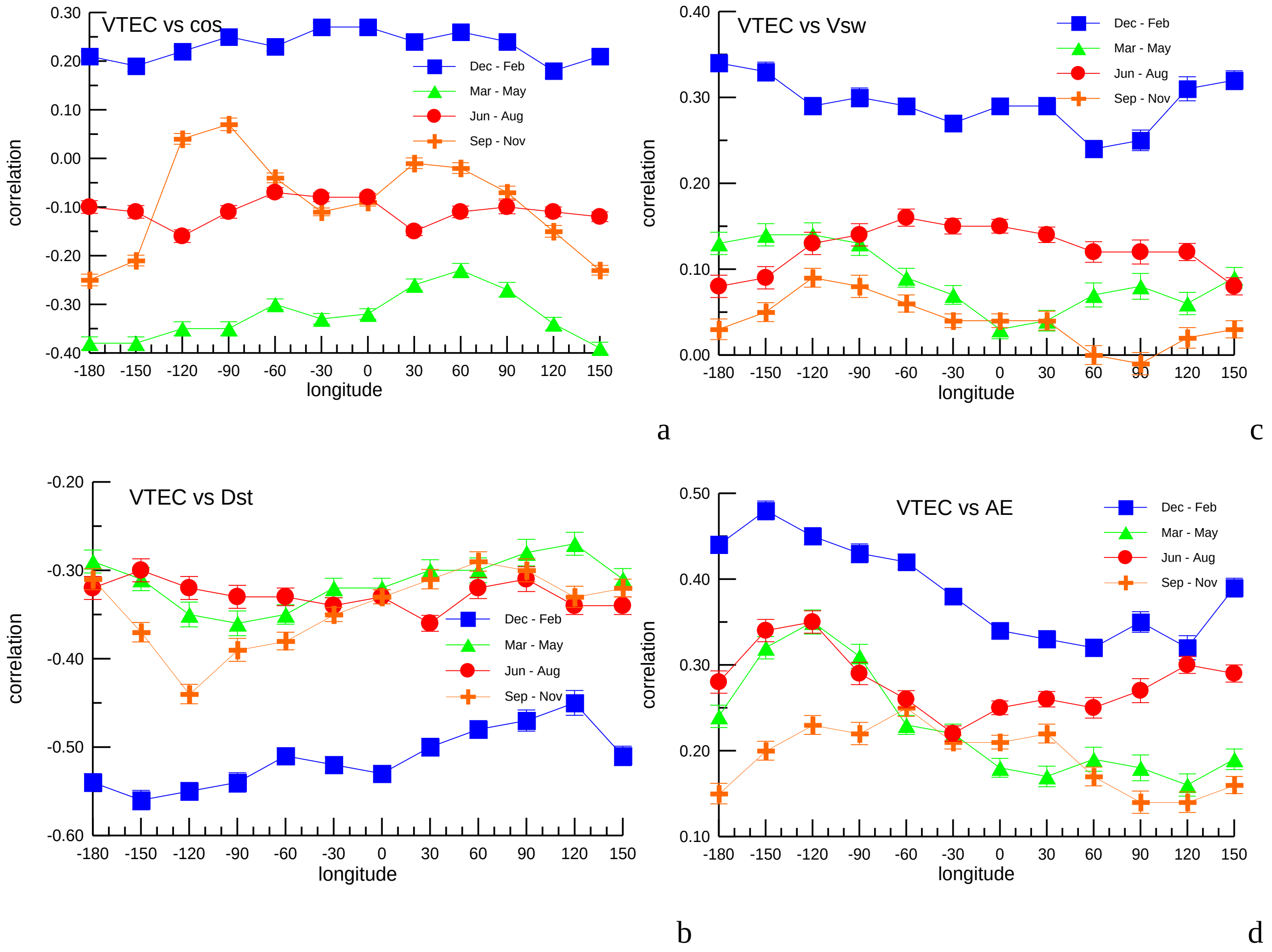

Figure 11. Seasonal variation of the correlation between the maxVTEC and (a) cosine of annual angle, (b) *Dst* index, (c) solar wind velocity, and (d) *AE* index. The cosine of annual angle shows a strong anti-correlation around spring equinox. High correlations with the solar wind and geomagnetic parameters are found during winter (Dec – Feb), especially in the Pacific and South America regions.

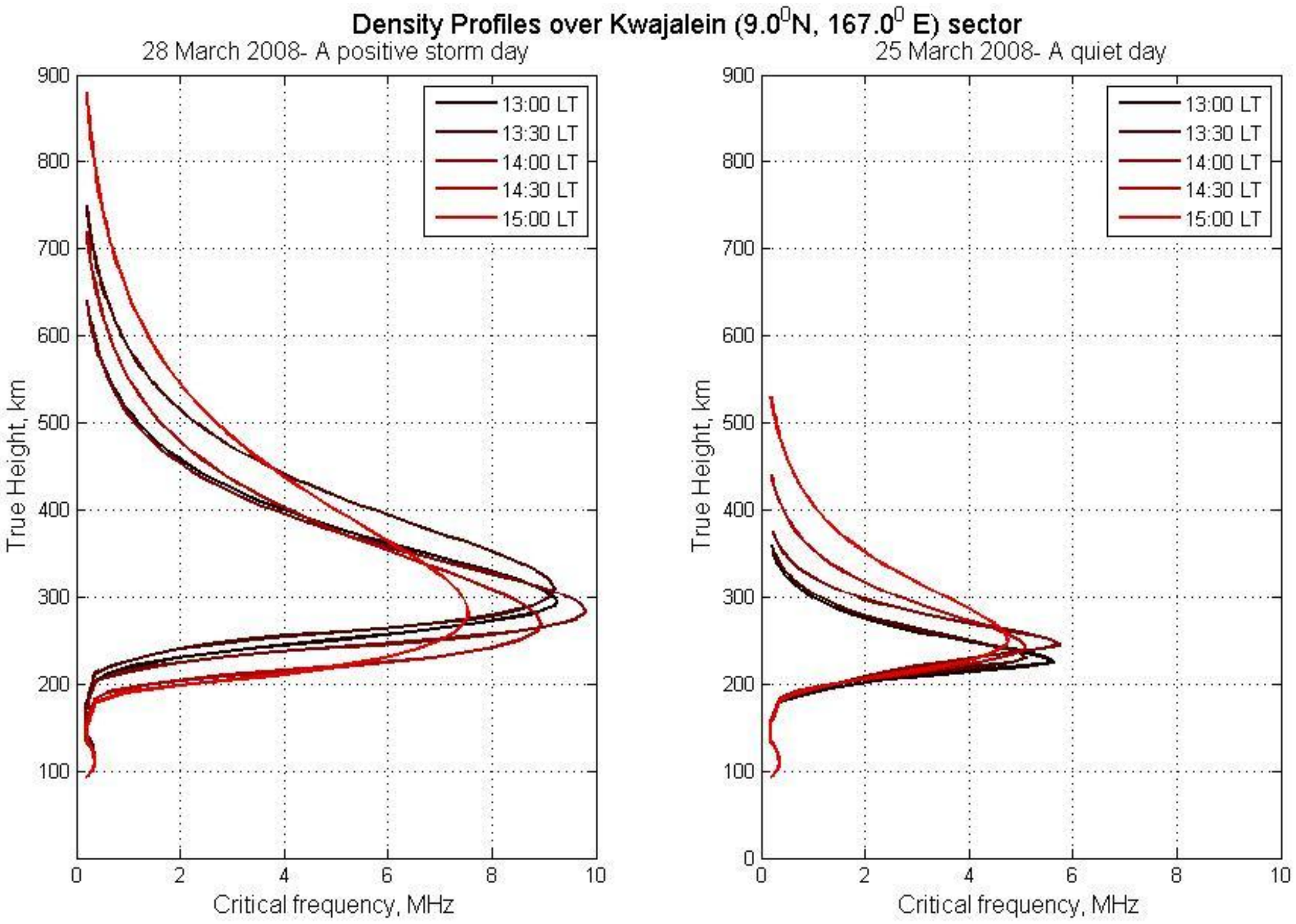

a

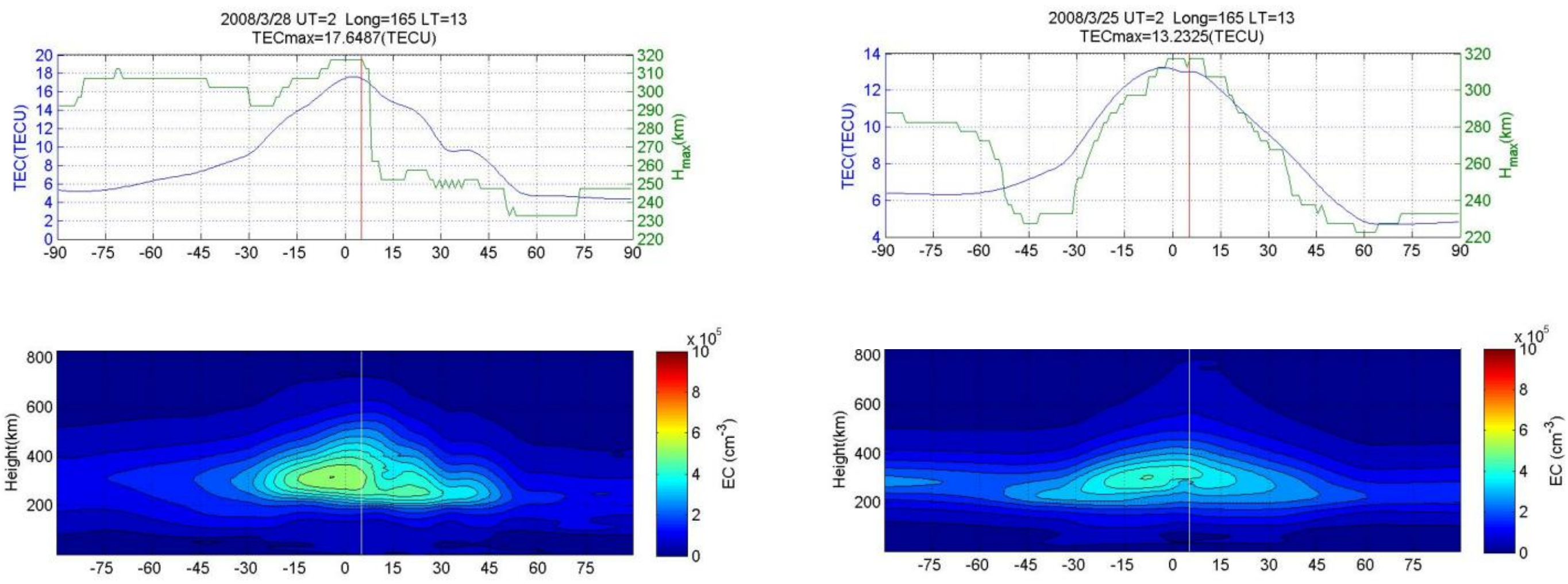

b

Figure 12. Height profile of electron content (EC) measured in the Pacific region (a) over Kwajalein (9°N, 167°E) and (b) reconstructed from COSMIC/FORMOSAT-3 radio-occultation tomography at longitude 167°E during a quiet day on 25 March 2008 (right panels) and in the maximum of RGS-related positive ionospheric storm on 28 March 2008 (left panels). During the storm, the EC in the Pacific region increases substantially and the maximum of F-layer elevates up to ~ 50 km from the height of ~250 km to ~300 km.

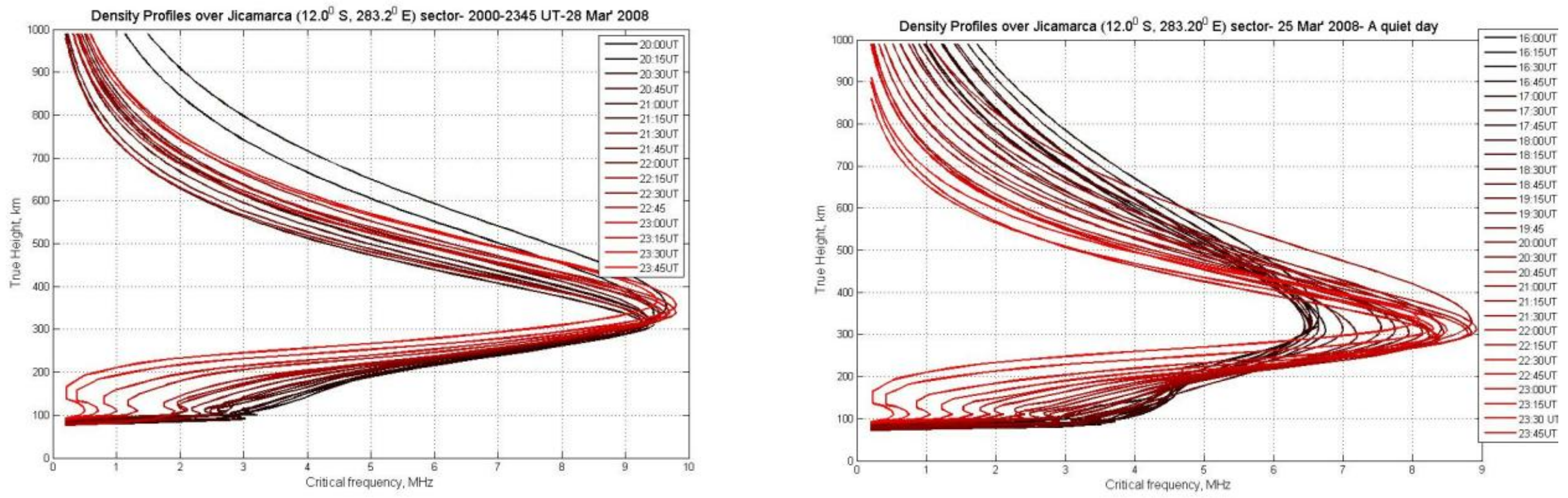

a

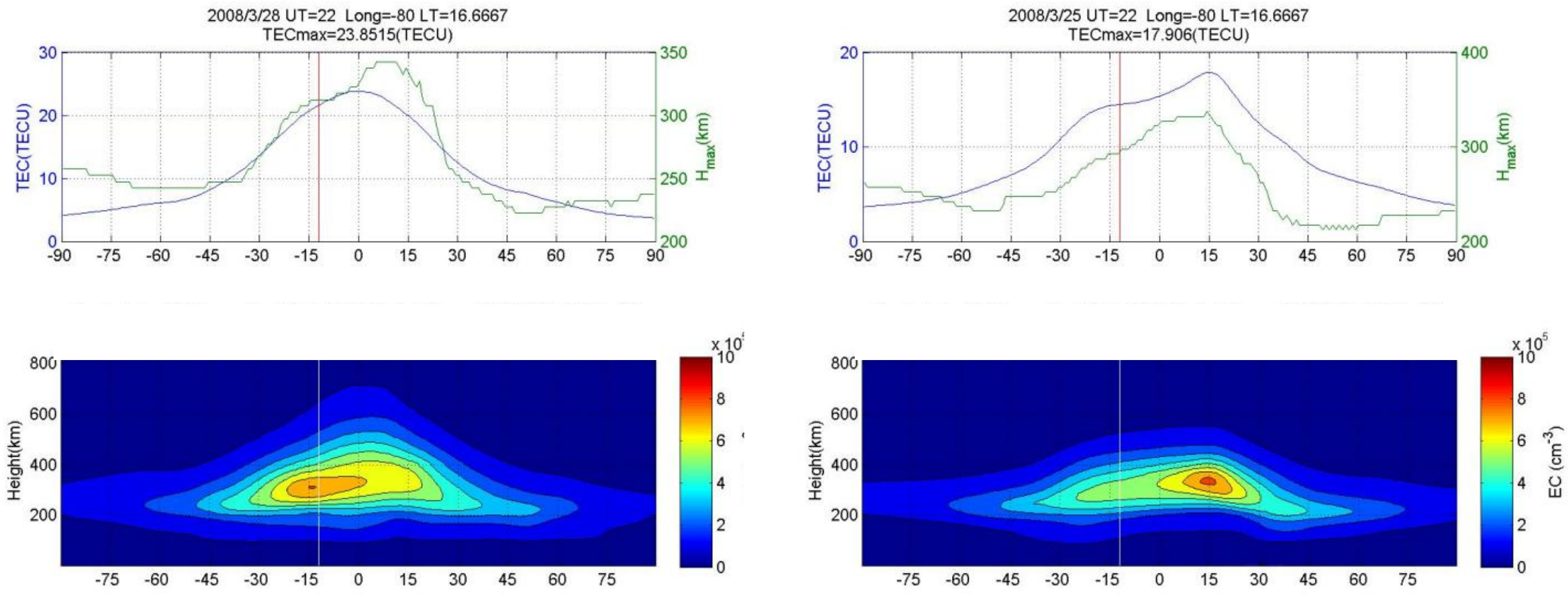

b

Figure 13. The same as in Figure 12, but over Jicamarka (12°S, 77°W) and at longitude 80°W. During the storm, the EC in the South America region increases substantially and the maximum of F-layer elevates up to ~ 50 km from the height of ~300 km to ~350 km.

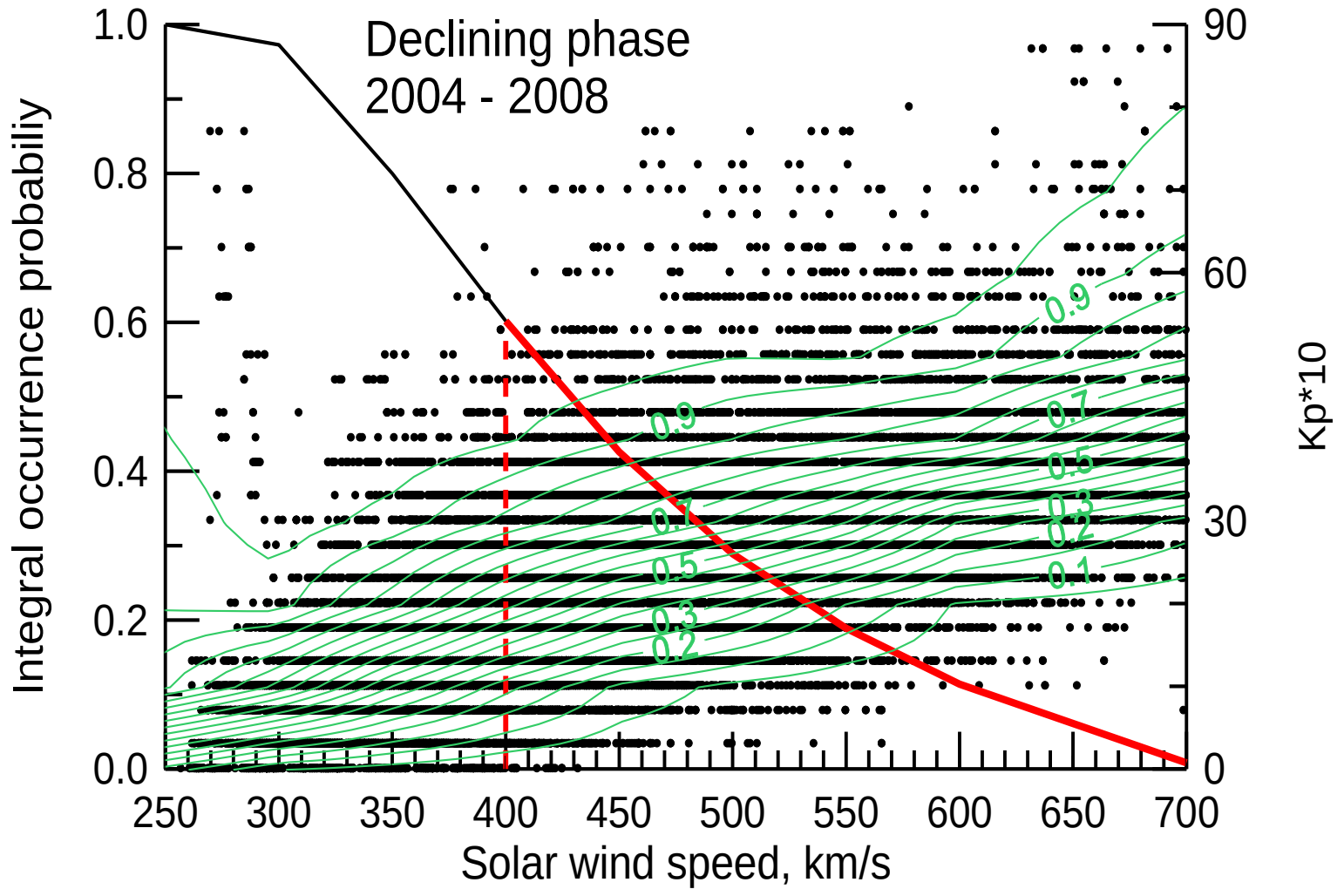

Figure 14. Integral probabilities: (solid curves, left axis) of occurrence of solar wind streams with velocities higher than given and (green isolines with numbers, right axis) of *Kp* smaller than given at various velocities in 2004 – 2008. Dots depict the scatter plot of *Kp* versus speed. High-speed solar wind streams with velocity >400 km/s (depicted by red curve) occur in 60% of statistics. More than 30% of statistics is characterized by small Kp < 3 under high-speed solar wind streams that is proper for RGSs.